\documentclass{iopjournal}

\usepackage[numbers,sort&compress]{natbib}

\usepackage{amsmath}
\usepackage{amssymb}
\usepackage{amsfonts}
\usepackage{graphicx}
\usepackage{dcolumn}
\usepackage{bm}

\usepackage[utf8]{inputenc}
\usepackage[T1]{fontenc}
\usepackage{mathptmx}
\usepackage{etoolbox}
\usepackage{tabularx}
\usepackage{makecell}
\usepackage{placeins}

\usepackage{enumitem}

\newcommand{\drho}{\delta\rho}

\newcommand{\dv}{\delta\bm{v}}

\begin{document}

\articletype{Paper} 

\title{Quantitative analysis of fluctuating hydrodynamics in uniform shear flow}

\author{Hiroyoshi Nakano$^{1*}$\orcid{0000-0000-0000-0000}, and Yuki Minami$^{2}$\orcid{0000-0000-0000-0000}}

\affil{$^1$Institute for Solid State Physics, University of Tokyo, 5-1-5, Kashiwanoha, Kashiwa 277-8581, Japan}

\affil{$^2$Faculty of Engineering, Gifu University, Yanagido, Gifu 501-1193, Japan}

\email{nakano.hiroyoshi.7n@issp.u-tokyo.ac.jp}


\begin{abstract}
Many theoretical predictions in fluctuating hydrodynamics under uniform shear flow have lacked precise quantitative verification because assessing the impact of analytical approximations is difficult and microscopic particle-based simulations have inherent limitations.
To address this problem, we perform direct numerical simulations of the fluctuating Navier-Stokes equations with shear-periodic boundary conditions.
We provide a decisive validation of two seminal frameworks: the Lutsko-Dufty theory for nonequilibrium long-range correlations, and the dynamic renormalization group (RG) theory pioneered by Forster, Nelson, and Stephen for anomalous transport.
First, we demonstrate that the predictions of the Lutsko-Dufty theory are quantitatively valid from the viscous-dominated, short-wavelength regime to the shear-dominated, long-wavelength regime.
Second, we test the quantitative predictive capability of the dynamic RG approach and show that the one-loop RG prediction is accurate even when the renormalization correction is comparable to the bare viscosity, a regime in which conventional perturbation theory fails.
Our findings solidify the foundations of these classical theories, paving the way for quantitative analyses using fluctuating hydrodynamics.
\end{abstract}

\section{Introduction}
\label{sec1}
Standard hydrodynamics serves as a successful macroscopic description of fluid motion, ranging from engineering applications to turbulent flows.
However, as the observation scale approaches the mesoscopic regime, thermal fluctuations of the constituent particles become significant, requiring a stochastic extension known as fluctuating hydrodynamics~\cite{Landau1987-eu}.
Within this framework, simple fluids are described by the fluctuating Navier-Stokes (NS) equations~\cite{Landau1987-eu, De_Zarate2006-xw, Das2011-ao}, which extend the deterministic NS equations with stochastic fluxes satisfying the fluctuation-dissipation theorem.
The scope of the fluctuating NS equations spans equilibrium fluctuations and nonequilibrium phenomena.
For instance, they describe generic long-range correlations under nonequilibrium conditions~\cite[and references therein]{De_Zarate2006-xw, Dorfman1994-cl, Bedeaux2015-lu, Sengers2024-gz}, as well as phenomena such as shear-induced nucleation~\cite{Furukawa2006-fx, Kurotani2020-zs} and interfacial instabilities at the nanoscale~\cite{Barker2023-ua}.

Historically, research on fluctuating hydrodynamics has been predominantly driven by theoretical approaches.
Since the 1970s, various analytical frameworks, such as linear approximations~\cite{Kawasaki1970-vo, Brogioli2001-ct, Wada2003-je, Peraud2017-xt}, mode-coupling theory (MCT)~\cite{Pomeau1975-ll, Das1986-rf, Spohn2014-ky, Nakano2025-ld}, and dynamic renormalization group (RG) analysis~\cite{Siggia1976-qg, Forster1977-lr, DeDominicis1979-dd, Kardar1986-td, Ertas1993-pd, Toner1995-qd, Canet2011-as, Gosteva2025-cq}, have been developed.
Beyond elucidating qualitative scaling laws---for example in dynamical critical phenomena~\cite{Hohenberg1977-zt} and the Kardar-Parisi-Zhang universality class~\cite{Halpin-Healy2015-ld, Takeuchi2018-xg}---these techniques have yielded explicit quantitative predictions.
Examples include prefactors for long-time tails~\cite{Kawasaki1970-vo}, predictions for nonequilibrium long-range correlations~\cite{Dorfman1994-cl, Vailati2011-rz, Takacs2011-gg, Nakano2022-kv, Srivastava2023-nx, Nakano2025-zw}, and universal constants in fully developed turbulence~\cite{Yakhot1986-mp}.
However, obtaining an explicit prediction does not by itself establish its quantitative accuracy.
These analytical treatments often rely on uncontrolled approximations whose quantitative effects are difficult to assess a priori, such as truncations of perturbative expansions in dynamic RG analysis.
Consequently, establishing the range of quantitative validity of these predictions remains a major challenge.

In this paper, we focus on the fluctuating NS equations under uniform shear flow to provide a quantitative verification of two seminal theories:
the Lutsko-Dufty theory for nonequilibrium long-range correlations in sheared fluids~\cite{Lutsko1985-zb, Lutsko1985-ib} and the dynamic RG theory for anomalous transport in two-dimensional (2D) fluids, pioneered by Forster, Nelson, and Stephen (FNS)~\cite{Forster1977-lr}.
These theories and their key approximations are summarized in Table~\ref{tab1}.

Previous studies have attempted to verify these specific predictions using lattice-gas cellular automata~\cite{Naitoh1990-nh, van-der-Hoef-MA1991-tj}, molecular dynamics simulations~\cite{Hoover1995-qy, Gravina1995-rt, Ferrario1997-ka, Bhattacharyya2000-jn, Isobe2008-xd, Choi2017-xt, Otsuki2009-kt, Otsuki2009-ld, Nakano2022-kv}, and multiparticle collision dynamics~\cite{Varghese2015-rk, Varghese2017-vw}.
However, isolating mesoscopic hydrodynamic behavior in microscopic simulations requires substantial computational resources and remains challenging.
Indeed, quantitative tests of the RG prediction using particle models remain inconclusive~\cite{Isobe2008-xd}, while low-wavenumber deviations from the Lutsko-Dufty prediction have been reported in microscopic simulations~\cite{Nakano2022-kv}.
It remains unclear whether these discrepancies arise from the analytical approximations, finite-size effects, or microscopic physics beyond the fluctuating-hydrodynamic description~\cite{Ortiz_de_Zarate2019-pp}.

To evaluate their quantitative performance, we employ direct numerical simulations (DNS) of the fluctuating NS equations.
Unlike particle-based methods, our DNS directly solve the fluctuating NS equations, allowing us to independently test the validity of the analytical approximations.
This approach is made possible by the steady advancement of numerical methodologies over the past two decades, particularly the highly accurate finite-volume schemes established by Bell, Donev, Garcia, and their collaborators~\cite{BalboaUsabiaga2012-sh, Delong2013-fh, Donev2014-jy, Donev2015-tm, Srivastava2023-nx, Garcia2024-nq}.
We extend their scheme to enforce shear-periodic boundary conditions (also known as Lees-Edwards boundary conditions), using a technique developed in the study of homogeneously sheared turbulence~\cite[and references therein]{Houssem-Kasbaoui2017-qy}.
While our previous work~\cite{Nakano2025-tj} demonstrated that the presence of solid walls introduces severe boundary effects that hinder the quantitative validation of bulk theories, the present implementation enables us to evaluate fluctuation effects in a bulk environment.

Using our DNS scheme, we first test the Lutsko-Dufty predictions beyond the regime assumed in their original derivation~\cite{Lutsko1985-zb}.
While their theory was originally formulated for the viscous-dominated regime at high wavenumbers, our numerical results show that its analytical expressions quantitatively reproduce the hydrodynamic fluctuations even in the shear-dominated and low-wavenumber regimes.
This finding provides a firm empirical foundation for the widespread use of the Lutsko-Dufty theory.
Second, we assess the quantitative reliability of the one-loop dynamic RG prediction for compressible fluids~\cite{Chen1995}.
By measuring the observed viscosity—a central macroscopic quantity in 2D anomalous transport—in our full nonlinear simulations, we find good quantitative agreement with this prediction well into the strongly renormalized regime.
This agreement extends into a regime where conventional perturbation theory is no longer quantitatively valid.

The remainder of this paper is organized as follows.
In Sec.~\ref{sec2}, we introduce the model.
In Sec.~\ref{sec3}, we detail the numerical scheme employed for our DNS.
In Sec.~\ref{sec4}, we perform a numerical verification of the Lutsko-Dufty theory by directly solving both the linearized and full nonlinear fluctuating NS equations.
In Sec.~\ref{sec5}, we quantitatively test the one-loop dynamic RG prediction by simulating the full nonlinear fluctuating NS equations.
In Sec.~\ref{sec6}, we justify applying this near-equilibrium RG prediction to sheared systems and derive an explicit criterion for the validity of this application.
Finally, Sec.~\ref{sec7} is devoted to concluding discussions.

\begin{table}[t]
\centering
\small 
\begin{tabularx}{1.\columnwidth}{|X|l|X|c|l|} 
\hline
\textbf{Target Phenomenon} & \textbf{Theoretical Basis} & \textbf{Key Approximations} & \textbf{Sec.} & \textbf{Ref.} \\ \hline
Long-range correlations & Linear theory & \makecell[l]{Linearization,\\ Mode decoupling approx.} & IV & \makecell[l]{Lutsko \& Dufty \\ (1985)~\cite{Lutsko1985-zb, Lutsko1985-ib}} \\ \hline
\makecell[l]{Nonlinear RG effects of \\ viscosity} & Dynamic RG theory & \makecell[l]{One-loop RG approx.} & V, VI & \makecell[l]{Forster, Nelson, \\ \& Stephen (1977)~\cite{Forster1977-lr} \\ Chen (1995)~\cite{Chen1995}} \\ \hline
\end{tabularx}
\caption{Overview of the theoretical frameworks examined quantitatively in this paper.}
\label{tab1}
\end{table}

\begin{figure}
\centering
\includegraphics[width=15cm]{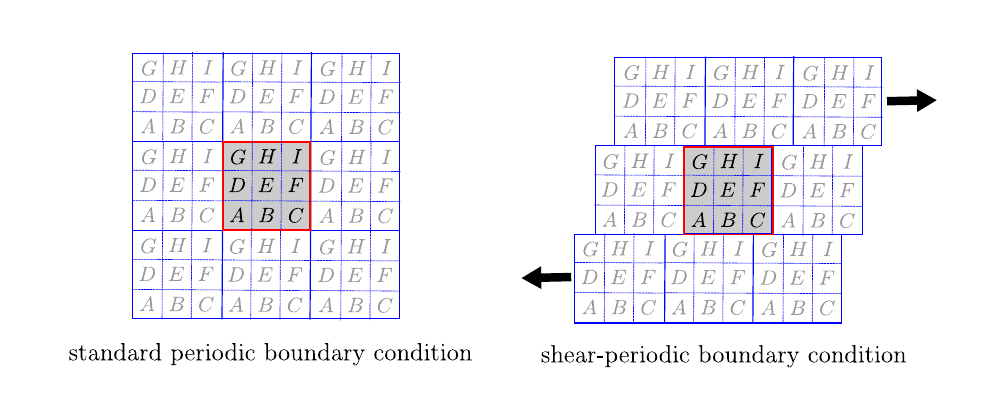}
\caption{\label{fig1}
Schematic illustrations comparing boundary conditions.
(Left) Standard periodic boundary conditions, where the central simulation box is surrounded by a lattice of stationary image boxes.
(Right) Shear-periodic (Lees-Edwards) boundary conditions, where the image boxes slide continuously with a relative velocity $\dot{\gamma} L_y$ in the flow direction.
This sliding motion induces uniform shear flow without introducing physical walls.
}
\end{figure}
\section{Model}
\label{sec2}
We consider a 2D compressible fluid at a constant temperature $T$.
The time evolution is governed by the fluctuating NS equations for the mass density field $\rho(\bm{r},t)$ and the velocity field $\bm{v}(\bm{r},t)$:
\begin{align}
    \frac{\partial \rho}{\partial t} &= - \nabla \cdot (\rho \bm{v}), \label{eq:sto_cont} \\
    \rho \left[ \frac{\partial \bm{v}}{\partial t} + (\bm{v} \cdot \nabla) \bm{v} \right] &= - \nabla p + \eta_0 \nabla^2 \bm{v} + \zeta_0 \nabla(\nabla \cdot \bm{v}) - \nabla \cdot \bm{\Pi}^{\mathrm{ran}}. \label{eq:sto_ns}
\end{align}
The pressure $p$ is determined by the isothermal equation of state, $p = c_T^2 \rho$, where $c_T$ is the isothermal sound speed.
Dissipation is characterized by the shear viscosity $\eta_0$ and the bulk viscosity $\zeta_0$.
The random stress tensor $\bm{\Pi}^{\mathrm{ran}}$ accounts for thermal fluctuations, modeled as Gaussian white noise satisfying the fluctuation-dissipation theorem:
\begin{align}
    \big\langle \Pi^{\mathrm{ran}}_{ij}(\bm{r},t)\Pi^{\mathrm{ran}}_{mn}(\bm{r}',t') \big\rangle = 2 k_B T \delta(\bm{r}-\bm{r}')\delta(t-t') \biggl[\eta_0 \left(\delta_{im}\delta_{jn}+\delta_{in}\delta_{jm}\right) + \left(\zeta_0 - \eta_0 \right) \delta_{ij}\delta_{mn} \biggr]. \label{eq:rand_st}
\end{align}

In this study, we investigate the nonequilibrium steady state under uniform shear flow.
Specifically, we consider a state characterized by a constant mean density and a linear mean velocity profile:
\begin{align}
    \rho(\bm{r}, t) = \rho_0, \qquad \bm{v}(\bm{r}, t) = \dot{\gamma} y \bm{e}_x,
    \label{eq:mean_field}
\end{align}
where $\dot{\gamma}$ is the shear rate and $\bm{e}_x$ is the unit vector in the $x$-direction.
The origin of the $y$ coordinate is chosen at the center of the simulation domain, so that $-L_y/2\leq y<L_y/2$ and the mean velocity vanishes at $y=0$.
To maintain this flow profile in a periodic domain and analyze bulk properties free from wall effects, we impose shear-periodic boundary conditions (also known as Lees-Edwards boundary conditions):
\begin{align}
    \begin{cases}
        \rho(x, y + L_y, t) = \rho(x - \dot{\gamma} L_y t, y, t), \\[3pt]
        v_x(x, y + L_y, t) = v_x(x - \dot{\gamma} L_y t, y, t) + \dot{\gamma} L_y, \\[3pt]
        v_y(x, y + L_y, t) = v_y(x - \dot{\gamma} L_y t, y, t).
    \end{cases}
    \label{eq:lebc}
\end{align}
Physically, these boundary conditions can be interpreted as an extension of standard periodic boundary conditions.
As illustrated in Fig.~\ref{fig1}, while standard periodic boundaries imply a lattice of stationary image boxes, shear-periodic boundaries correspond to image boxes sliding relative to the central box with a velocity $\dot{\gamma} L_y$, thereby sustaining uniform shear flow.
We note that introducing solid walls induces confinement effects that severely complicate the analysis of bulk properties; for a detailed discussion on such wall effects, we refer the reader to our previous work~\cite{Nakano2025-tj}.

With the mean flow profile fixed by Eq.~\eqref{eq:mean_field}, our primary focus lies in the fluctuations around this steady state.
To this end, we define the velocity fluctuations $\delta\bm{v}(\bm{r}, t)$ as
\begin{align}
    \delta\bm{v}(\bm{r}, t) &= \bm{v}(\bm{r}, t) - \dot{\gamma} y \bm{e}_x. \label{eq:v_fluct}
\end{align}
The density fluctuation is simply defined as
\begin{align}
    \delta\rho = \rho - \rho_0. \label{eq:rho_fluct}
\end{align}

\begin{figure}
\centering
\includegraphics[width=5cm]{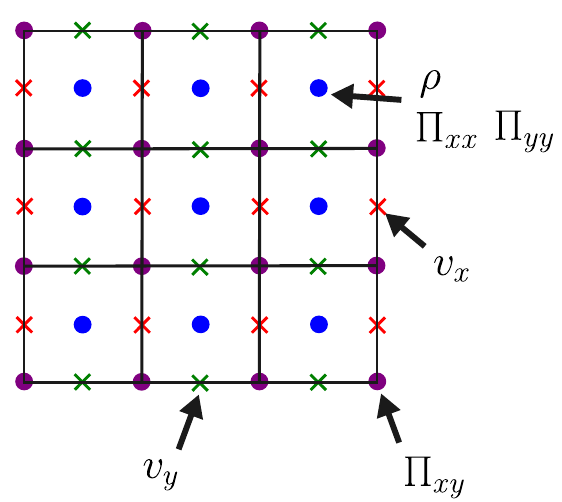}
\caption{\label{fig2} 
Schematic representation of the staggered grid layout used for spatial discretization. Different field variables are discretized at distinct locations on the grid.
}
\end{figure}
\section{Implementation of the Numerical Simulations}
\label{sec3}
This section presents the numerical scheme used to solve the fluctuating hydrodynamic equations.
In the simulation, we solve the evolution equations for the density $\rho$ and the momentum density fluctuation $\delta \bm{j} := \rho \delta \bm{v}$:
\begin{align}
    & \left(\frac{\partial}{\partial t} + \dot{\gamma} y \frac{\partial}{\partial x} \right) \rho = - \nabla \cdot \delta \bm{j}, \label{eq:sto_cont_sh} \\[5pt]
    & \left(\frac{\partial}{\partial t} + \dot{\gamma} y \frac{\partial}{\partial x} \right) \delta\bm{j} + \dot{\gamma} \delta j_y\bm{e}_x + \nabla \cdot (\delta\bm{j}\delta\bm{v}) = - \nabla p + \eta_0 \nabla^2 \delta\bm{v} + \zeta_0 \nabla(\nabla \cdot \delta\bm{v}) - \nabla \cdot \bm{\Pi}^{\mathrm{ran}}. \label{eq:sto_ns_sh}
\end{align}

To integrate these equations, we combine the high-accuracy spatial discretization scheme of Srivastava et al.~\cite{Srivastava2023-nx} with a shear-handling technique developed for homogeneously sheared turbulence~\cite{Houssem-Kasbaoui2017-qy}.
Our study represents the first application of this shear-handling technique to fluctuating hydrodynamics, which enables a precise evaluation of hydrodynamic fluctuations without introducing numerical dissipation under shear-periodic boundary conditions.

\subsection{Spatial Discretization}
\label{sec3-1}
We discretize Eqs.~\eqref{eq:sto_cont_sh} and \eqref{eq:sto_ns_sh} in real space on a staggered grid.
As illustrated in Fig.~\ref{fig2}, physical quantities are defined at distinct locations on the grid cells:
\begin{itemize}
\item \textbf{Cell Centers:} Scalar quantities ($\rho, p$) and the diagonal components of the stochastic stress tensor ($\Pi^{\mathrm{ran}}_{xx}, \Pi^{\mathrm{ran}}_{yy}$).
\item \textbf{Cell Faces:} Vector quantities ($j_x, v_x$ at vertical faces; $j_y, v_y$ at horizontal faces).
\item \textbf{Cell Edges:} The off-diagonal stress component ($\Pi^{\mathrm{ran}}_{xy}$).
\end{itemize}
Spatial derivatives are evaluated using second-order central differences, while the interpolation of variables to different grid locations is performed via linear averaging.
For details, we refer the reader to the original works by Balboa Usabiaga et al.~\cite{BalboaUsabiaga2012-sh} and Srivastava et al.~\cite{Srivastava2023-nx}, as well as our previous study~\cite{Nakano2025-tj}.

Throughout this paper, we set the grid spacing to be uniform in both directions, $\Delta x = \Delta y = h$.
The system size is set to $L_x = L_y (:= L) = N h$, where $N$ is the number of grid cells in one direction.

\subsection{Time integration}
\label{sec3-2}
For the time integration, we adapt an operator-splitting algorithm based on the method developed by Houssem Kasbaoui et al.~\cite{Houssem-Kasbaoui2017-qy}.
Specifically, to advance the system from time step $n$ to $n+1$ over a time interval $\Delta t$, we split the governing equations Eqs.~\eqref{eq:sto_cont_sh} and \eqref{eq:sto_ns_sh} into two sequential stages: a non-advective step and an advection step.

\subsubsection*{Non-advective Step}
In the first stage, we integrate the equations of motion excluding the mean shear advection terms ($\dot{\gamma} y \partial_x$):
\begin{equation}
    \left\{
    \begin{aligned}
        & \frac{\partial \rho}{\partial t} = - \nabla \cdot \delta \bm{j}, \\
        & \frac{\partial \delta\bm{j}}{\partial t} + \dot{\gamma} \delta j_y \bm{e}_x + \nabla \cdot (\delta\bm{j}\delta\bm{v}) = - \nabla p + \eta_0 \nabla^2 \delta\bm{v} + \zeta_0 \nabla(\nabla \cdot \delta\bm{v}) - \nabla \cdot \bm{\Pi}^{\mathrm{ran}}.
    \end{aligned}
    \right.
    \label{eq:sto_eq_p1}
\end{equation}
The time integration of this system is performed using the low-storage three-stage Runge-Kutta method, as employed by Delong et al.~\cite{Delong2013-fh}.

\subsubsection*{Advection Step}
Following the non-advective step, we apply advection by the mean shear flow.
The governing equations for this step are purely advective:
\begin{equation}
    \left\{
    \begin{aligned}
        & \frac{\partial \rho}{\partial t} + \dot{\gamma} y \frac{\partial \rho}{\partial x} = 0, \\
        & \frac{\partial \delta\bm{j}}{\partial t} + \dot{\gamma} y \frac{\partial \delta\bm{j}}{\partial x} = 0.
    \end{aligned}
    \right.
    \label{eq:sto_eq_p2}
\end{equation}
Let us denote any of the advected fields ($\rho, \delta j_x$, or $\delta j_y$) generically as $\phi$.
If $\phi$ were a spatially continuous field, the exact solution to Eq.~\eqref{eq:sto_eq_p2} over a time step $\Delta t$ would be a simple spatial translation along the $x$-direction:
\begin{align}
    \phi(x, y, t + \Delta t) = \phi(x - \dot{\gamma} y \Delta t, y, t).
    \label{eq:shift_continuous_field}
\end{align}
However, in our simulation, the field $\phi$ is discretized on the staggered grid, which we denote as $(x_i, y_j)$ (the discrete locations depend on the type of the variable $\phi$, as illustrated in Fig.~\ref{fig2}).
Evaluating the shifted value $\phi(x - \dot{\gamma} y \Delta t, y, t)$ requires some form of spatial interpolation, because the continuous displacement $\dot{\gamma} y_j \Delta t$ is generally not an integer multiple of the grid spacing.
We then employ a discrete Fourier interpolation method, an approach proven to be highly accurate in the DNS of homogeneously sheared turbulence~\cite{Houssem-Kasbaoui2017-qy}.

In practice, this method evaluates the shift by expanding the field into a 1D discrete Fourier series along the $x$-axis:
\begin{align}
    \phi(x_i, y_j, t) = \sum_{k_x} \tilde{\phi}(k_x, y_j, t) e^{i k_x x_i},
\end{align}
where $\tilde{\phi}$ are the discrete Fourier coefficients and $k_x$ are the discrete wavenumbers.
By applying the exact continuous displacement within this Fourier representation, the advected field at $t + \Delta t$ is accurately reconstructed as:
\begin{align}
    \phi(x_i, y_j, t + \Delta t) &= \phi(x_i - \dot{\gamma} y_j \Delta t, y_j, t) \nonumber \\
    &= \sum_{k_x} \tilde{\phi}(k_x, y_j, t) e^{i k_x (x_i - \dot{\gamma} y_j \Delta t)}.
\end{align}
In our implementation, this operation is performed using the fast Fourier transform (FFT).
Specifically, at each fixed $y_j$, we apply a one-dimensional FFT along the $x$ direction, multiply each Fourier coefficient by the corresponding phase factor, and then apply the inverse FFT.
This operation gives the exact solution of the shear-advection substep for all retained Fourier modes and therefore introduces no interpolation-induced numerical diffusion.
However, we note that the sequential splitting of the non-advective and shear-advection substeps introduces a temporal discretization error, making the overall time integration first-order accurate in $\Delta t$.

\section{Nonequilibrium Long-Range Correlations}
\label{sec4}
In this section, we focus on the static correlation function of velocity fluctuations under uniform shear flow and provide a quantitative verification of the theoretical predictions by Lutsko and Dufty~\cite{Lutsko1985-zb}.

\subsection{Static Correlation Functions}
\label{sec4-1}
We consider the equal-time correlation matrix of the velocity fluctuations in Fourier space.
The Fourier transform of the velocity fluctuations $\delta\bm{v}(\bm{r},t)$ is defined as $\delta\tilde{\bm{v}}(\bm{k},t) = \int d\bm{r}\, \delta\bm{v}(\bm{r},t) e^{-i\bm{k}\cdot\bm{r}}$, and we define the static correlation matrix $C_{\alpha\beta}(\bm{k})$ as:
\begin{align}
    \langle \delta\tilde{v}_{\alpha}(\bm{k}, t) \delta\tilde{v}_{\beta}^*(\bm{k}', t) \rangle = V \delta_{\bm{k},\bm{k}'} C_{\alpha\beta}(\bm{k}),
    \label{eq:corr_definition}
\end{align}
where the asterisk $\ast$ represents the complex conjugate, $V := L^2$ is the area of the system, and the indices $\alpha, \beta$ denote the Cartesian components $\{x, y\}$.

In thermal equilibrium, the equipartition theorem dictates that these correlations are purely local and diagonal:
\begin{align}
    C_{\alpha\beta}^{\mathrm{eq}}(\bm{k}) = \delta_{\alpha\beta} \frac{k_B T}{\rho_0}.
\end{align}
In contrast, for nonequilibrium steady states, it is well known that long-range correlations generically emerge, characterized by an algebraic power-law dependence on $\bm{k}$.
Lutsko and Dufty calculated this explicit dependence for sheared fluids by analyzing the linearized fluctuating hydrodynamic equations.

\subsection{Theoretical Predictions in Lutsko and Dufty (1985)}
\label{sec4-2}
First, we review the theoretical predictions of Lutsko and Dufty~\cite{Lutsko1985-zb}~\footnote{
While the original theory addressed a three-dimensional fluid with energy conservation, the mathematical structure is fundamentally similar to the two-dimensional isothermal system considered here. We therefore derive the corresponding analytical predictions for our specific model to allow for direct comparison.}.
Following their theoretical framework, we analyze Eqs.~\eqref{eq:sto_cont} and \eqref{eq:sto_ns} within the linear approximation.
Neglecting the nonlinear terms leads to the equations for the fluctuations:
\begin{align}
    \left(\frac{\partial}{\partial t} + \dot{\gamma} y \frac{\partial}{\partial x}\right) \delta\rho &= - \rho_0 \nabla \cdot \delta\bm{v}, \\
    \left(\frac{\partial}{\partial t} + \dot{\gamma} y \frac{\partial}{\partial x}\right) \delta\bm{v} + \dot{\gamma} \delta v_y \bm{e}_x &= - \frac{c_T^2}{\rho_0}\nabla \delta\rho + \nu_0 \nabla^2 \delta\bm{v} + \nu_B \nabla(\nabla \cdot \delta\bm{v}) + \bm{f},
\end{align}
where $\bm{f} = -\rho_0^{-1}\nabla \cdot \bm{\Pi}^{\mathrm{ran}}$ is the random force vector, $\nu_0 = \eta_0/\rho_0$ is the kinematic shear viscosity, and $\nu_B = \zeta_0/\rho_0$ is the kinematic bulk viscosity.

These equations are analyzed in Fourier space.
To simplify the problem, it is advantageous to decompose the velocity fluctuations into longitudinal ($L$) and transverse ($T$) eigenmodes.
Using the projection matrix, the velocity fluctuations are transformed as:
\begin{align}
    \begin{pmatrix}
        \delta\tilde{v}_L(\bm{k}) \\
        \delta\tilde{v}_T(\bm{k})
    \end{pmatrix}
    =
    \begin{pmatrix}
        \hat{k}_x & \hat{k}_y \\
        -\hat{k}_y & \hat{k}_x
    \end{pmatrix}
    \begin{pmatrix}
        \delta\tilde{v}_x(\bm{k}) \\
        \delta\tilde{v}_y(\bm{k})
    \end{pmatrix}.
    \label{eq:projection}
\end{align}
where $\hat{k}_\alpha \equiv k_\alpha/k$.
The correlations in the Cartesian basis are related to these eigenmode correlations by:
\begin{subequations}
    \begin{align}
        C_{xx}(\bm{k}) &= \hat{k}_x^2 C_{LL}(\bm{k}) + \hat{k}_y^2 C_{TT}(\bm{k}) - 2 \hat{k}_x \hat{k}_y \mathrm{Re}[C_{LT}(\bm{k})], \\
        C_{yy}(\bm{k}) &= \hat{k}_y^2 C_{LL}(\bm{k}) + \hat{k}_x^2 C_{TT}(\bm{k}) + 2 \hat{k}_x \hat{k}_y \mathrm{Re}[C_{LT}(\bm{k})], \\
        C_{xy}(\bm{k}) &= \hat{k}_x \hat{k}_y (C_{LL}(\bm{k}) - C_{TT}(\bm{k})) + (\hat{k}_x^2 - \hat{k}_y^2) \mathrm{Re}[C_{LT}(\bm{k})].
    \end{align}
\label{eq:C_decomp_all}
\end{subequations}
Here, the invariance of the uniform shear state under spatial inversion ensures that $C_{LT}(\bm{k})$ is real.

To derive an analytical solution, Lutsko and Dufty restricted their analysis to the small-$k$ region and the viscous-dominated regime, where viscous damping outweighs shear advection (i.e., $\dot{\gamma} \lesssim \nu_0 k^2$).
Under these conditions, they introduced a mode-decoupling approximation and simplified the complex coupling between density and velocity fluctuations under shear.
This procedure consists of two steps:

\paragraph{(i) Decoupling of $L$ and $T$ modes:} The coupling between the transverse mode $\delta \tilde{v}_T$ and the longitudinal mode $\delta \tilde{v}_L$ is found to be of higher order in $\bm{k}$. In the small-$\bm{k}$ limit, these modes can be treated as statistically independent, allowing us to neglect their cross-correlation:
\begin{align}
    C_{LT}(\bm{k}) \approx 0.
\label{eq:CLT_zero}
\end{align}
\paragraph{(ii) Simplification of the coupling between the $L$ and $\rho$ modes:} The remaining coupling between the density $\delta\rho$ and the longitudinal velocity $\delta v_L$ is calculated by a perturbative expansion up to $\mathcal{O}(k^2)$.

Under these approximations, the linearized equations become analytically tractable, yielding the following explicit forms for the static correlation functions:
\begin{align}
    C_{TT}(\bm{k}) &= \frac{k_B T}{\rho_0} + \dot{\gamma} \frac{2 k_B T}{\rho_0} \frac{k_x}{k^2} \int_0^{\infty} ds \, (k_y + \dot{\gamma} s k_x) e^{-2\nu_0 I(\bm{k},s)}, \label{eq:CTT_theory} \\
    C_{LL}(\bm{k}) &= \frac{k_B T}{\rho_0} - \dot{\gamma} \frac{k_B T}{\rho_0} k k_x \int_0^{\infty} ds \frac{k_y + \dot{\gamma}sk_x}{[k_x^2 + (k_y + \dot{\gamma} s k_x)^2]^{3/2}} e^{-\Gamma_0 I(\bm{k},s)}, \label{eq:CLL_theory}
\end{align}
where $I(\bm{k},s)$ is given by
\begin{align}
    I(\bm{k},s) &= s k^2 + \dot{\gamma} s^2 k_x k_y + \frac{1}{3} \dot{\gamma}^2 s^3 k_x^2,
\end{align}
and $\Gamma_0 := (\eta_0 + \zeta_0)/\rho_0$ is the longitudinal damping coefficient.
These expressions have been extensively used in the literature~\cite{Wada2003-je, De_Zarate2006-xw, Ortiz_de_Zarate2019-pp}.

\begin{figure}
\centering
\includegraphics{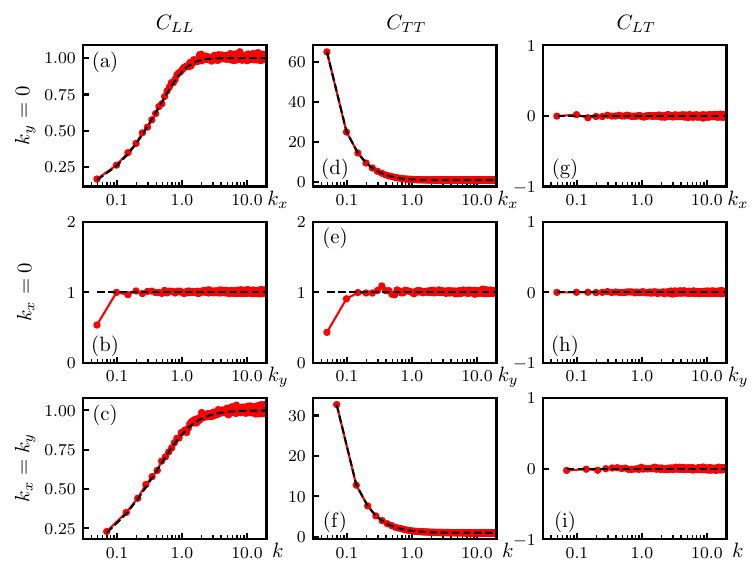}
\caption{\label{fig3} 
Simulation results of the velocity correlations in the linearized fluctuating NS equations.
(a-c) Longitudinal correlation $C_{LL}(\bm{k})$, (d-f) Transverse correlation $C_{TT}(\bm{k})$, and (g-i) Cross-correlation $C_{LT}(\bm{k})$.
The red symbols represent the numerical data, while the black dashed lines correspond to the theoretical predictions given by Eqs.~\eqref{eq:CTT_theory} and \eqref{eq:CLL_theory}.
The theoretical prediction for $C_{LT}(\bm{k})$ is identically zero due to the decoupling approximation.
Parameters: $N=768$, $h = 1/6$, $\dot{\gamma} = 0.1$, $\eta_0 = \zeta_0 = 0.1$, $\rho_0 = k_B T = 1.0$, and $c_T^2 = 5000$.
}
\end{figure}
\subsection{Numerical Verification}
\label{sec4-3}

We now verify the theoretical predictions in Eqs.~\eqref{eq:CTT_theory} and \eqref{eq:CLL_theory} by comparing them with our numerical simulation results.
The simulations are performed by numerically solving the linearized fluctuating NS equations (i.e., neglecting the nonlinear advection terms), and the results are obtained by statistical averaging in the nonequilibrium steady state.
Given that the Lutsko-Dufty theory involves linearized approximations, our DNS provide a direct and independent test of the mathematical simplifications employed in the Lutsko-Dufty derivation, most notably the mode-decoupling approximation.
See Appendix~\ref{secA} for details on the averaging procedures.

Figure~\ref{fig3} presents the static correlation functions for the longitudinal mode ($C_{LL}$, left column), the transverse mode ($C_{TT}$, middle column), and the cross-correlation ($C_{LT}$, right column).
To elucidate the strong anisotropy induced by the shear flow, we plot the correlations along three distinct cuts in Fourier space: the flow direction ($k_y=0$), the gradient direction ($k_x=0$), and the diagonal direction ($k_x=k_y$).

As shown in this figure, the analytical expressions derived by Lutsko and Dufty accurately reproduce the DNS data for all three correlation functions across all plotted directions and the entire wavenumber range.
Crucially, the cross-correlation $C_{LT}(\bm{k})$ vanishes within statistical error (right column) across the entire wavenumber range investigated, which validates the mode-decoupling approximation [Eq.~\eqref{eq:CLT_zero}].
As further support, we provide two-dimensional heat maps of the correlations in Appendix~\ref{secB}.

We clarify the range over which this agreement holds.
As detailed in Appendix~\ref{secC}, the Lutsko-Dufty expressions show that the velocity correlations under uniform shear exhibit a crossover between the viscous-dominated and shear-dominated regimes.
The corresponding transverse and longitudinal crossover wavenumbers are
\begin{align}
    k_c^T \sim \sqrt{\frac{\dot{\gamma}}{\nu_0}},
    \label{eq:k_cross_T} \\
    k_c^L \sim \sqrt{\frac{\dot{\gamma}}{\Gamma_0}}.
    \label{eq:k_cross_L}
\end{align}
The range $|\bm{k}|\gtrsim k_c^{T,L}$ is the viscous-dominated regime. 
This is the regime in which Lutsko and Dufty originally justified the mode-decoupling approximation.
In this regime, the leading nonequilibrium contributions take the following forms:
\begin{align}
    C_{TT}^{\rm lin}(\bm{k})-\frac{k_BT}{\rho_0}
    &\simeq
    \frac{k_BT}{\rho_0}
    \frac{\dot{\gamma}}{\nu_0}
    \hat{k}_x\hat{k}_y
    k^{-2},
    \nonumber \\
    \frac{k_BT}{\rho_0}-C_{LL}^{\rm lin}(\bm{k})
    &\simeq
    \frac{k_BT}{\rho_0}
    \frac{\dot{\gamma}}{\Gamma_0}
    \hat{k}_x\hat{k}_y
    k^{-2}.
    \label{eq:ld_viscous_scaling_main}
\end{align}
The range $|\bm{k}|\lesssim k_c^{T,L}$ is the shear-dominated regime, where shear advection changes the leading small-$k$ behavior to
\begin{align}
    C_{TT}^{\rm lin}(\bm{k})-\frac{k_BT}{\rho_0}
    &\simeq
    1.183\,
    \frac{k_BT}{\rho_0}
    \left(\frac{\dot{\gamma}}{\nu_0}\right)^{2/3}
    |\hat{k}_x|^{2/3}
    k^{-4/3},
    \nonumber \\
    \frac{k_BT}{\rho_0} - C_{LL}^{\rm lin}(\bm{k})
    &\simeq
    \frac{k_BT}{\rho_0} - 0.939\,
    \frac{k_BT}{\rho_0}
    \left(\frac{\dot{\gamma}}{\Gamma_0}\right)^{-1/3}
    |\hat{k}_x|^{-1/3}
    k^{2/3},
    \label{eq:ld_shear_scaling_main}
\end{align}
The derivation of these expressions is detailed in Appendix~\ref{secC}.

\begin{figure}
\centering
\includegraphics{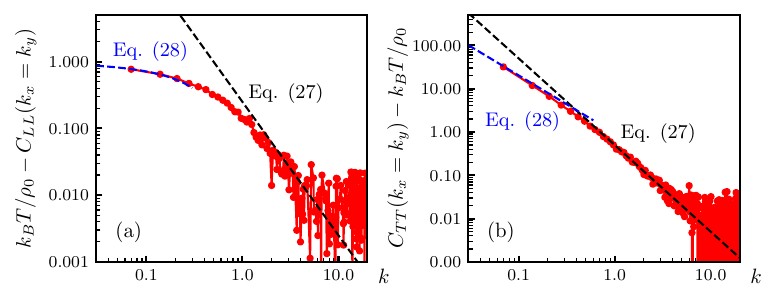}
\caption{\label{fig4}
Log--log plots of the shear-induced deviations from equilibrium in the DNS data.
The DNS data are the same as those used in Figs.~\ref{fig3}(c) and (f).
(a) Longitudinal contribution $k_BT/\rho_0-C_{LL}^{\rm lin}(\bm{k})$.
(b) Transverse contribution $C_{TT}^{\rm lin}(\bm{k})-k_BT/\rho_0$.
The red symbols represent the DNS data.
The black and blue dashed lines represent the viscous-dominated and shear-dominated predictions in Eqs.~\eqref{eq:ld_viscous_scaling_main} and \eqref{eq:ld_shear_scaling_main}, respectively.
Their amplitudes are determined by the simulation parameters, without fitting to the DNS data.
}
\end{figure}

Figure~\ref{fig4} replots the DNS data in Figs.~\ref{fig3}(c) and (f), showing the shear-induced contributions from equilibrium on log--log scales.
The black and blue dashed lines represent the viscous-dominated and shear-dominated predictions in Eqs.~\eqref{eq:ld_viscous_scaling_main} and \eqref{eq:ld_shear_scaling_main}, respectively.
The DNS data agree quantitatively with both asymptotic predictions, including their amplitudes, in the corresponding wavenumber ranges.
From this agreement, we confirm that the Lutsko-Dufty expressions remain quantitatively accurate across the crossover between the two regimes.
In particular, the mode-decoupling approximation remains accurate in the shear-dominated regime, although it was originally justified only in the viscous-dominated regime.

\FloatBarrier
\subsection{Full Nonlinear Simulation}
\label{sec4-4}

\begin{figure}
\centering
\includegraphics{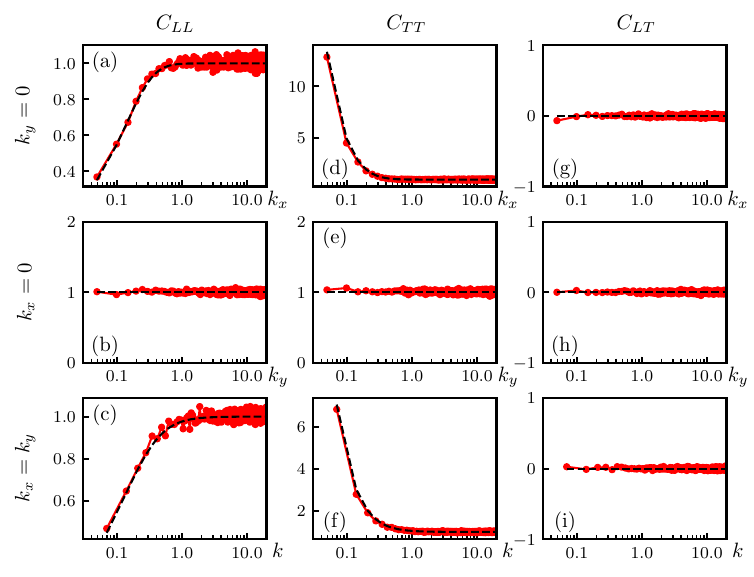}
\caption{\label{fig5}
Simulation results of the velocity correlations in the full nonlinear fluctuating NS equations.
The panel layout, plotted correlations, and theoretical curves are the same as in Fig.~\ref{fig3}; the red symbols represent the nonlinear DNS data.
Parameters: $N=768$, $h=1/6$, $\dot{\gamma}=0.1$, $\eta_0=\zeta_0=1.0$, $\rho_0=k_B T=1.0$, and $c_T^2=5000$.
}
\end{figure}

We further test the Lutsko-Dufty expressions by solving the full nonlinear fluctuating NS equations in a large-viscosity regime, where nonlinear effects are weak.
Figure~\ref{fig5} shows that the nonlinear DNS agrees with the Lutsko-Dufty predictions within statistical accuracy, and the cross-correlation $C_{LT}(\bm{k})$ remains negligible.
Thus, the Lutsko-Dufty theory retains quantitative predictive power even for the full nonlinear equations when the viscosity is chosen appropriately.
In Sec.~\ref{sec5}, we quantify the nonlinear contribution and thereby identify the parameter range corresponding to this weakly nonlinear regime.

\section{Nonlinear Contributions to the Observed Viscosity}
\label{sec5}
In this section, we focus on the macroscopic transport properties of the 2D fluid.
Dynamic RG provides a systematic framework for describing how nonlinear interactions among hydrodynamic fluctuations modify transport coefficients across different length scales.
The application of this framework to the fluctuating NS equations was pioneered by FNS~\cite{Forster1977-lr}, who predicted anomalous transport in two dimensions.

In practice, the RG flow equations are commonly truncated at the one-loop level.
Our aim in this section is to quantitatively assess the resulting one-loop RG prediction by comparing it with simulations of the full nonlinear fluctuating NS equations, in which nonlinear advection is retained without perturbative truncation.
Because our simulations describe a compressible fluid, we evaluate this prediction using the compressible formulation~\cite{Chen1995}, which includes both transverse and longitudinal hydrodynamic modes.
The original incompressible FNS result
\footnote{
FNS considered three noise models. The noise in the present compressible formulation is the counterpart of their Model A.
}
is recovered when the longitudinal contribution is suppressed (see also Appendix~\ref{secF}).

\subsection{Bare and Observed Viscosities}
\label{sec5-1}
We first define the macroscopic shear viscosity calculated from observables, which serves as a key quantity to detect the nonlinear effects predicted by the RG analysis.
By rewriting the fluctuating NS equations [Eqs.~\eqref{eq:sto_cont} and \eqref{eq:sto_ns}] in the form of a momentum continuity equation, the momentum flux tensor $\Pi_{ab}$ for a two-dimensional compressible fluid is defined as
\begin{align}
    \Pi_{ab} = \rho v_a v_b + p \delta_{ab} - \eta_0 (\partial_a v_b + \partial_b v_a - \delta_{ab} \nabla \cdot \bm{v}) - \zeta_0 \delta_{ab} \nabla \cdot \bm{v}.
\end{align}
In a steady state under uniform shear flow, the macroscopic shear stress is given by $\sigma_{xy} = -\langle \Pi_{xy} \rangle$.
Using the velocity fluctuations, this can be explicitly written as
\begin{align}
    \sigma_{xy} = \eta_0 \dot{\gamma} - \langle \rho \delta v_x \delta v_y \rangle.   
    \label{eq:def_stress}
\end{align}
Thus, the effective shear viscosity calculated from the observables is given by
\begin{align}
    \eta_{\rm obs} \equiv \frac{\sigma_{xy}}{\dot{\gamma}} = \eta_0 - \frac{\langle \rho \delta v_x \delta v_y \rangle}{\dot{\gamma}}.
    \label{eq:def_eta_obs}
\end{align}
Here, $\eta_0$ is the \textit{bare shear viscosity} appearing in the fluctuating NS equations.
It cannot be directly detected through bulk fluid measurements because the observation of macroscopic shear stress intrinsically includes the correction from thermal fluctuations~\cite{Nakano2025-tj}.
In contrast, $\eta_{\rm obs}$ is the experimentally tractable quantity; therefore, we refer to this quantity as the \textit{observed shear viscosity}.
Furthermore, the difference $\Delta\eta \equiv \eta_{\rm obs} - \eta_0$ is termed the mode-coupling contribution~\cite{Pomeau1975-ll, Lutsko1985-ib} or the renormalization correction~\cite{Forster1977-lr}.

A fundamental characteristic of two-dimensional fluids is that the observed viscosity $\eta_{\rm obs}$ diverges with system size due to the infrared (IR) divergence of the mode-coupling contribution.
This phenomenon, known as low-dimensional anomalous transport, signifies the breakdown of classical macroscopic transport laws.
Dynamic RG captures the asymptotic scale dependence of this divergence through scale-dependent transport coefficients.

\subsection{One-Loop Dynamic RG Prediction}
\label{sec5-2}

We summarize the one-loop dynamic RG prediction used to calculate the observed viscosity.
The central concept of the RG analysis is coarse-graining.
This procedure iteratively eliminates velocity fluctuations with short wavelengths (high-wavenumber modes in the shell $\Lambda e^{-l} < k < \Lambda$) and rescales the equations to derive an effective description for the long-wavelength physics.
Through this analysis, the nonlinear advection term in the fluctuating NS equations vanishes asymptotically in the macroscopic limit in two and higher dimensions.
Instead, the nonlinear interactions among the eliminated fast modes act as an effective dissipation mechanism and renormalize the viscosity coefficient.
As a result, in the hydrodynamic limit ($k \to 0$), the effect of nonlinear advection can be absorbed into scale-dependent transport coefficients.

For a compressible fluid, the bare shear and bulk viscosities, $\eta_0$ and $\zeta_0$, are replaced by the renormalized viscosities $\eta_{\rm ren}(k)$ and $\zeta_{\rm ren}(k)$:
\begin{align}
    \frac{\partial \delta \tilde{\rho}}{\partial t}
    &=
    - i\rho_0 \bm{k}\cdot \delta\tilde{\bm{v}}, 
\label{eq:renor_cont} \\
    \rho_0 \frac{\partial \delta\tilde{\bm{v}}}{\partial t}
    &=
    -i\bm{k} c_T^2 \delta\tilde{\rho}
    -\eta_{\rm ren}(k) k^2 \delta\tilde{\bm{v}}
    -\zeta_{\rm ren}(k)\bm{k}(\bm{k}\cdot\delta\tilde{\bm{v}})
    + \tilde{\bm{F}}^{\rm ran}.
\label{eq:renor_ns}
\end{align}
Here, $k=|\bm{k}|$.
At the one-loop level, the long-wavelength behavior of these renormalized viscosities is governed by the following RG flow equations~\cite{Chen1995}:
\begin{align}
    \frac{d\eta_{\rm ren}(k)}{d\log k} &= -\frac{k_B T\rho_0}{16\pi}
    \left[ \frac{1}{\eta_{\rm ren}(k)} + \frac{1}{G_{\rm ren}(k)} \right],
    \label{eq:chen_rg_eta} \\
    \frac{dG_{\rm ren}(k)}{d\log k} &= -\frac{3k_B T\rho_0}{16\pi}
    \left[ \frac{1}{\eta_{\rm ren}(k)} + \frac{1}{G_{\rm ren}(k)} \right],
    \label{eq:chen_rg_G}
\end{align}
where
\begin{align}
    G_{\rm ren}(k) \equiv \eta_{\rm ren}(k)+\zeta_{\rm ren}(k).
\end{align}

The incompressible FNS result is recovered when density fluctuations are suppressed,
$\delta\rho=0$, and the velocity field is restricted to the transverse sector,
$\bm{k}\cdot\delta\tilde{\bm{v}}=0$.
In this limit, the longitudinal contribution in Eq.~\eqref{eq:chen_rg_eta} is absent, and the RG equation reduces to
\begin{align}
    \frac{d\eta_{\rm ren}(k)}{d\log k}
    =
    -\frac{k_B T\rho_0}{16\pi}
    \frac{1}{\eta_{\rm ren}(k)}.
\end{align}
With the condition $\eta_{\rm ren}(k_{\rm uv})=\eta_0$ imposed at the UV cutoff $k_{\rm uv}=2\pi/a_{\rm uv}$, the solution of this equation is
\begin{align}
    \eta_{\rm ren}(k)
    =
    \sqrt{
    \eta_0^2
    +
    \frac{k_B T \rho_0}{8\pi}
    \log\left(\frac{k_{\rm uv}}{k}\right)
    }.
    \label{eq:fns_viscosity}
\end{align}
This is the standard incompressible FNS result for Model A, and the same expression is also obtained from MCT~\cite{Pomeau1975-ll}.
See also Appendix~\ref{secF} for a discussion of the incompressible limit in fluctuating hydrodynamics.

For finite compressibility, the longitudinal contribution must be retained, and Eqs.~\eqref{eq:chen_rg_eta} and \eqref{eq:chen_rg_G} do not reduce to the simple square-root form of the incompressible FNS theory.
In the comparison below, we solve these coupled RG equations numerically subject to the conditions
\begin{align}
    \eta_{\rm ren}(k_{\rm uv})=\eta_0,
    \qquad
    G_{\rm ren}(k_{\rm uv})=\eta_0+\zeta_0,
\end{align}

We now apply these near-equilibrium RG results to our nonequilibrium shear flow setup and relate the renormalized viscosity to the observed viscosity.
The RG equations, Eqs.~\eqref{eq:chen_rg_eta} and \eqref{eq:chen_rg_G}, were obtained in the absence of a mean shear.
Therefore, the use of the resulting renormalized viscosities $\eta_{\rm ren}(k)$ and $\zeta_{\rm ren}(k)$ is restricted to the regime in which the shear-induced distortion of the fluctuations is perturbatively small.
As we will show in Sec.~\ref{sec6}, this condition corresponds to the low-Reynolds-number regime, $Re \equiv \dot{\gamma}L^2/\nu_0 \ll 1$.
Under this condition, the effect of uniform shear is incorporated by adding the shear-advection terms to the effective equations, Eqs.~\eqref{eq:renor_cont} and \eqref{eq:renor_ns}:
\begin{align}
    \left(\frac{\partial}{\partial t} - \dot{\gamma} k_x \frac{\partial}{\partial k_y}\right)
    \delta \tilde{\rho}(\bm{k},t)
    &= - i\rho_0 \bm{k}\cdot \delta\tilde{\bm{v}}(\bm{k},t),
    \nonumber \\
    \rho_0 \left(\frac{\partial}{\partial t} - \dot{\gamma} k_x \frac{\partial}{\partial k_y}\right)
    \delta \tilde{\bm{v}}(\bm{k},t)
    + \rho_0 \dot{\gamma}\, \delta \tilde{v}_y(\bm{k},t)\bm{e}_x
    &= - i\bm{k}c_T^2\delta\tilde{\rho}(\bm{k},t)
    - \eta_{\rm ren}(k) k^2 \delta\tilde{\bm{v}}(\bm{k},t) \nonumber \\
    &\quad
    - \zeta_{\rm ren}(k)\bm{k}\left[\bm{k}\cdot\delta\tilde{\bm{v}}(\bm{k},t)\right]
    + \tilde{\bm{F}}^{\rm ran}(\bm{k},t).
\end{align}
Following the Lutsko-Dufty treatment, we can derive analytical expressions for the steady-state velocity correlations from these equations, denoted by $C_{LL}^{\rm eff}(\bm{k})$, $C_{TT}^{\rm eff}(\bm{k})$, and $C_{LT}^{\rm eff}(\bm{k})$.
Within the same mode-decoupling approximation, $C_{LT}^{\rm eff}\simeq 0$.
Substitution into Eq.~\eqref{eq:def_eta_obs}, together with the decomposition in Eq.~\eqref{eq:C_decomp_all}, gives
\begin{align}
    \eta_{\rm obs}
    &= \eta_0
    - \frac{\rho_0}{\dot{\gamma}}
    \int_{k_{\rm IR}}^{k_{\rm uv}} \frac{d^2\bm{k}}{(2\pi)^2}
    \hat{k}_x\hat{k}_y C_{LL}^{\rm eff}(\bm{k})
    + \frac{\rho_0}{\dot{\gamma}}
    \int_{k_{\rm IR}}^{k_{\rm uv}} \frac{d^2\bm{k}}{(2\pi)^2}
    \hat{k}_x\hat{k}_y C_{TT}^{\rm eff}(\bm{k}),
\end{align}
where $k_{\rm IR}=2\pi/L$.
In the low-Reynolds-number regime, the Lutsko-Dufty solution of the effective equations reduces to the viscous-dominated asymptotic forms given in Appendix~\ref{secD2}:
\begin{align}
    C_{TT}^{\rm eff}(\bm{k})
    &\simeq
    \frac{k_B T}{\rho_0}
    + \dot{\gamma}\,\frac{k_B T}{\eta_{\rm ren}(k)}
    \frac{k_xk_y}{k^4},
    \nonumber \\
    C_{LL}^{\rm eff}(\bm{k})
    &\simeq
    \frac{k_B T}{\rho_0}
    - \dot{\gamma}\,\frac{k_B T}{G_{\rm ren}(k)}
    \frac{k_xk_y}{k^4}.
\end{align}
Substitution of these asymptotic forms gives
\begin{align}
    \eta_{\rm obs}
    &\approx \eta_0
    + k_B T\rho_0
    \int_{k_{\rm IR}}^{k_{\rm uv}} \frac{d^2\bm{k}}{(2\pi)^2}
    \frac{k_x^2k_y^2}{k^6}
    \left[
        \frac{1}{\eta_{\rm ren}(k)}
        +
        \frac{1}{G_{\rm ren}(k)}
    \right]
    \nonumber \\
    &= \eta_0
    + \frac{k_B T\rho_0}{16\pi}
    \int_{k_{\rm IR}}^{k_{\rm uv}} \frac{dk}{k}
    \left[
        \frac{1}{\eta_{\rm ren}(k)}
        +
        \frac{1}{G_{\rm ren}(k)}
    \right]
    \nonumber \\
    &= \eta_{\rm ren}\left(k=\frac{2\pi}{L}\right).
    \label{eq:obs_eta_fns}
\end{align}
Further details of the calculation leading to Eq.~\eqref{eq:obs_eta_fns} are given in Appendix~\ref{secD3}.
Thus, in the low-Reynolds-number regime, the observed viscosity is identified with the renormalized shear viscosity evaluated at the infrared cutoff set by the system size.

It is instructive to examine the weak-renormalization limit.
In this regime, the renormalized viscosities in Eq.~\eqref{eq:obs_eta_fns} can be replaced by their bare values, yielding
\begin{align}
    \eta_{\rm obs}
    &\approx \eta_0
    + \frac{k_B T\rho_0}{16\pi}
    \left(
        \frac{1}{\eta_0}
        +
        \frac{1}{\eta_0+\zeta_0}
    \right)
    \log \left(\frac{L}{a_{\rm uv}}\right).
    \label{eq:obs_eta_pert}
\end{align}
This expression is mathematically identical to the prediction derived at the lowest order of a simple perturbative expansion (see Sec.~\ref{sec6}).
It also shows explicitly that the viscosity correction in the compressible formulation consists of the transverse contribution proportional to $1/\eta_0$ and the longitudinal contribution proportional to $1/(\eta_0+\zeta_0)$.

\subsection{Numerical Verification}
\label{sec5-3}
We now test the validity of the theoretical prediction in Eq.~\eqref{eq:obs_eta_fns} by comparing it with DNS of the fluctuating NS equations.
By simulating the full nonlinear dynamics, our DNS captures nonlinear effects to all orders, thereby enabling a direct quantitative test of the one-loop RG prediction.

\subsubsection{Soft UV Cutoff}

For a quantitative comparison between the DNS and the RG prediction, we must specify how the UV regularization introduced by the numerical grid is represented in the continuum theory.
In the RG formulation, the UV scale is introduced as a sharp circular cutoff, $k_{\rm uv}=2\pi/a_{\rm uv}$, whereas the numerical discretization may not produce such a sharp cutoff.

In particular, the staggered-grid interpolation introduces smooth wave-number-dependent factors.
For example, interpolation of the $x$ component of the velocity from the cell faces to the cell centers gives
\begin{align}
    (\overline{\delta v_x})_{i,j}
    =
    \sum_{\bm{k}}
    \cos\left(\frac{k_xh}{2}\right)
    \delta\tilde{v}_x(\bm{k})
    e^{i\bm{k}\cdot\bm{r}_{i,j}}.
\end{align}
The cosine factor gradually suppresses short-wavelength fluctuations instead of eliminating all modes at a single wave number.
Thus, the numerical regularization should be regarded as a soft UV cutoff.

\begin{figure}
\centering
\includegraphics{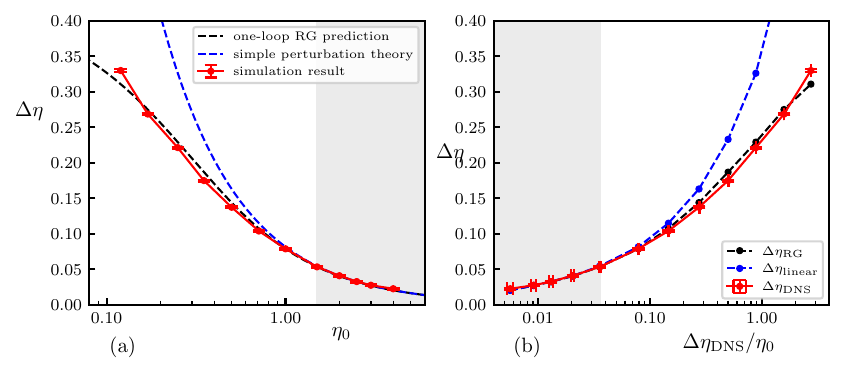}
\caption{\label{fig6}
The renormalization correction to the observed viscosity, $\Delta\eta=\eta_{\rm obs}-\eta_0$, in the low-Reynolds-number regime.
(a) $\Delta\eta$ is plotted as a function of the bare viscosity $\eta_0$.
(b) The DNS, RG, and perturbative values of $\Delta\eta$ are plotted as functions of the dimensionless ratio $\Delta\eta_{\rm DNS}/\eta_0$, which quantifies the strength of the nonlinear renormalization.
The red symbols represent the numerical results obtained from the full nonlinear fluctuating NS simulations.
The black dashed line represents the one-loop RG prediction given by Eq.~\eqref{eq:obs_eta_fns}.
The blue dashed line represents its weak-renormalization limit, which is equivalent to the lowest-order perturbative prediction in Eq.~\eqref{eq:obs_eta_pert}.
The light-gray regions indicate the data range used to fit
$a_{\rm uv}^{\rm eff}$.
Parameters: $N=96$, $h=1/6$, $\dot{\gamma}=0.02$, $\rho_0=k_B T=1$, $c_T^2=5000$, $\zeta_0=4\eta_0$, and $a_{\rm uv}^{\rm eff}=0.526$.
}
\end{figure}

In our analysis, we approximate this soft regularization by an effective sharp cutoff length, $a_{\rm uv}^{\rm eff}$, which we determine through a fitting procedure.
To isolate the numerical UV regularization from the nonlinear renormalization effects, we fit $a_{\rm uv}^{\rm eff}$ using only data satisfying
\begin{align}
    \eta_0\geq1.5.
\end{align}
These data lie in the weak-renormalization regime,
$\Delta\eta_{\rm DNS}/\eta_0<0.1$,
where the RG prediction reduces to the perturbative expression in Eq.~\eqref{eq:obs_eta_pert}.
For the data used in Fig.~\ref{fig6}, for which $\zeta_0=4\eta_0$, this procedure gives
\begin{align}
    a_{\rm uv}^{\rm eff}=0.526\pm0.028.
\end{align}

As shown in Appendix~\ref{secE}, we apply the same fitting procedure independently to the additional data at $\zeta_0/\eta_0=1$ and $19$.
The three independently fitted values of $a_{\rm uv}^{\rm eff}$ agree within about $2\%$.
This agreement provides strong evidence that the UV regularization is controlled primarily by the numerical lattice structure rather than by the fluid parameters, and supports our use of an effective sharp cutoff to represent the numerical UV regularization.

\subsubsection{Main Results}

Figure~\ref{fig6}(a) presents the dependence of the viscosity correction $\Delta\eta\equiv\eta_{\rm obs}-\eta_0$ on the bare shear viscosity $\eta_0$.
The simulations were performed by varying $\eta_0$ from $0.12$ to $4.0$ while fixing the system size and shear rate.
The bulk viscosity was varied simultaneously according to $\zeta_0=4\eta_0$.
To test the quantitative predictive capability of the one-loop RG theory, we compare the DNS results with the lowest-order perturbative prediction in Eq.~\eqref{eq:obs_eta_pert} and the one-loop RG prediction in Eq.~\eqref{eq:obs_eta_fns}.

The results in Fig.~\ref{fig6}(a) show that, at large $\eta_0$, where the renormalization correction is weak, the perturbative and RG predictions are nearly identical and both reproduce the DNS data.
This agreement is expected because the DNS data in this regime are used to determine $a_{\rm uv}^{\rm eff}$.
As $\eta_0$ decreases, however, the perturbative prediction deviates from the DNS data, whereas the RG prediction continues to describe them accurately.
This highlights the advantage of the one-loop RG approach over the lowest-order perturbative approximation in the strongly renormalized regime.

To directly assess the range of validity of the RG prediction, Fig.~\ref{fig6}(b) replots the same results against $\Delta\eta_{\rm DNS}/\eta_0$, which provides a normalized measure of the renormalization correction.
From this figure, we find that for $\zeta_0=4\eta_0$, the agreement between the RG prediction and the DNS data persists up to the largest $\Delta\eta_{\rm DNS}/\eta_0$ examined here, $\Delta\eta_{\rm DNS}/\eta_0\simeq3$.
As shown in Appendix~\ref{secE}, simulations with $\zeta_0/\eta_0=1$ and $19$ lead to the same general conclusion.
Considering all three values of $\zeta_0/\eta_0$, we obtain the following criterion for the quantitative reliability of the one-loop RG prediction:
\begin{align}
    \frac{\Delta\eta_{\rm DNS}}{\eta_0}\lesssim1.
\end{align}
This is the main result of this section.
For comparison, we note that the lowest-order perturbative prediction is reliable only in the weak-renormalization regime,
\begin{align}
    \frac{\Delta\eta_{\rm DNS}}{\eta_0}\lesssim0.1.
\end{align}
As shown in Sec.~\ref{sec6}, this lowest-order perturbative prediction is obtained using the Lutsko-Dufty correlations.
Therefore, this condition provides a practical criterion for the weakly nonlinear regime in which the Lutsko-Dufty theory is expected to remain quantitatively reliable (Sec.~\ref{sec4}).

We note that, at stronger renormalization, $\Delta\eta_{\rm DNS}/\eta_0\gtrsim1$, the deviation becomes more dependent on the value of $\zeta_0/\eta_0$.
A possible explanation for this dependence is the difference between the wave-number domains employed in the continuum RG calculation and the DNS.
The RG transformation eliminates modes in circular wave-number shells, whereas the discrete Fourier modes of the numerical grid occupy the square domain
$0<\max(|k_x|,|k_y|)\leq\pi/h$.
In our procedure, this difference in domain geometry is represented only through the fitted value of $a_{\rm uv}^{\rm eff}$, so its effects cannot be eliminated completely.
In addition, the numerical modes near the grid scale are weighted by the lattice interpolation.
Consequently, small residual discrepancies between the continuum RG prediction and the DNS results are expected even after $a_{\rm uv}^{\rm eff}$ has been determined.

\section{Shear-Rate Dependence of Observed Viscosity}
\label{sec6}
We finally address the shear-rate dependence of the observed viscosity.
The RG prediction reviewed in the previous section relies on a near-equilibrium treatment of hydrodynamic fluctuations and is therefore expected to break down under strong uniform shear.
In this section, we identify the regime in which the near-equilibrium RG prediction remains valid.
The discussion below follows Ref.~\cite{Ortiz_de_Zarate2019-pp}.

\subsection{Leading-Order Contribution of Nonlinear Advection}
\label{sec6-1}
We employ a simple perturbative approach to capture the leading-order shear-induced behavior of $\eta_{\rm obs}$.
We introduce a formal expansion parameter $\epsilon$ into the compressible fluctuating NS equation Eq.~\eqref{eq:sto_ns}:
\begin{align}
    \rho \left[ \frac{\partial \bm{v}}{\partial t} + \epsilon (\bm{v} \cdot \nabla) \bm{v} \right] = - \nabla p + \eta_0 \nabla^2 \bm{v} + \zeta_0 \nabla(\nabla \cdot \bm{v}) - \nabla \cdot \bm{\Pi}^{\mathrm{ran}}.
\end{align}
The fluctuation fields are expanded in powers of $\epsilon$ as $\delta \psi = \delta \psi^{(0)} + \epsilon \delta \psi^{(1)} + \cdots$, where $\psi$ denotes either $\drho$ or $\dv$.
The zeroth-order solutions, $\drho^{(0)}$ and $\dv^{(0)}$, are simply the solutions to the linearized hydrodynamics discussed in Sec.~\ref{sec4}.
Substituting these expansions into the definition of $\eta_{\rm obs}$, Eq.~\eqref{eq:def_eta_obs}, we find that the leading-order correction to $\eta_0$ is given by the correlation of the linearized fields, $\langle \delta v_x^{(0)} \delta v_y^{(0)} \rangle$, leading to the following explicit form for $\eta_{\rm obs}$:
\begin{align}
    \eta_{\rm obs} &\approx \eta_0 - \frac{\rho_0}{\dot{\gamma}} \langle \delta v_x^{(0)} \delta v_y^{(0)} \rangle \nonumber \\
    &= \eta_0 - \frac{\rho_0}{\dot{\gamma} V} \sum_{\bm{k}} \hat{k}_x \hat{k}_y C_{LL}^{\rm lin}(\bm{k}) + \frac{\rho_0}{\dot{\gamma} V} \sum_{\bm{k}} \hat{k}_x \hat{k}_y C_{TT}^{\rm lin}(\bm{k}),
    \label{eq:eta_oneloop}
\end{align}
where $C_{LL}^{\rm lin}(\bm{k})$ and $C_{TT}^{\rm lin}(\bm{k})$ are the static correlation functions in the Lutsko-Dufty theory.
Equivalent expressions employing the same perturbative approximation have been used in previous theoretical studies~\cite{Lutsko1985-ib, Wada2003-je, Ortiz_de_Zarate2019-pp}.

\begin{figure}
\centering
\includegraphics{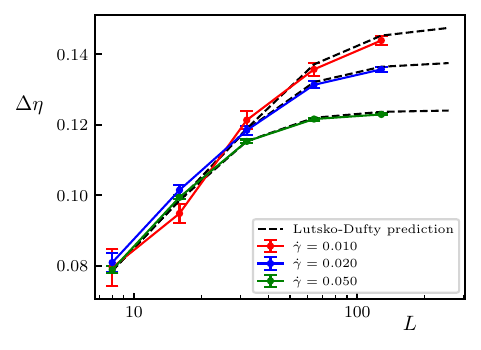}
\caption{\label{fig7} 
System-size dependence of the renormalization correction to the observed viscosity, $\Delta \eta = \eta_{\rm obs} - \eta_0$, for three different shear rates $\dot{\gamma}=0.01$ (red), $\dot{\gamma}=0.02$ (blue), and $\dot{\gamma}=0.05$ (green).
The colored symbols represent the simulation results, while the black dashed lines indicate the theoretical predictions obtained by substituting the Lutsko-Dufty expressions [Eqs.~\eqref{eq:CTT_theory} and \eqref{eq:CLL_theory}] into Eq.~\eqref{eq:eta_oneloop}.
The effective cutoff $a_{\rm uv}^{\rm eff}=0.725$ is determined by fitting the $\dot{\gamma}=0.020$ data and is then held fixed for the other two shear rates.
Parameters: $h=1/6$, $\rho_0=k_B T=1.0$, $c_T^2=5000$, and $\eta_0=\zeta_0=1.0$.
}
\end{figure}

Figure~\ref{fig7} presents the simulation results for the system-size dependence of the viscosity correction, $\Delta \eta \equiv \eta_{\rm obs} - \eta_0$.
Here, $\Delta \eta$ is evaluated via Eq.~\eqref{eq:eta_oneloop} using the velocity correlations $\langle \delta v_x^{(0)} \delta v_y^{(0)} \rangle$ directly measured from the DNS of the linearized fluctuating NS equations.
The data are plotted for three distinct shear rates: $\dot{\gamma} = 0.010$ (red), $0.020$ (blue), and $0.050$ (green).

For the theoretical curves, we substitute the Lutsko-Dufty expressions [Eqs.~\eqref{eq:CTT_theory} and \eqref{eq:CLL_theory}] into Eq.~\eqref{eq:eta_oneloop} and evaluate the discrete wave-number sum over the square domain $0<\max(|k_x|,|k_y|)\leq 2\pi/a_{\rm uv}^{\rm eff}$.
The effective cutoff determined in Sec.~\ref{sec5-3} cannot be directly transferred to the present calculation, because continuum RG transformations are conventionally formulated in the continuous circular domain $0<|\bm{k}|\leq 2\pi/a_{\rm uv}^{\rm eff}$ and the analysis in Sec.~\ref{sec5-3} likewise uses this circular domain.

We therefore determine $a_{\rm uv}^{\rm eff}$ separately for Fig.~\ref{fig7} by fitting all system sizes at $\dot{\gamma}=0.020$, obtaining $a_{\rm uv}^{\rm eff}=0.725$.
The same value is then used for $\dot{\gamma}=0.010$ and $0.050$ without further fitting.
The resulting curves show good quantitative agreement with the simulation results over the full range of system sizes.

This figure clearly demonstrates how this leading-order nonlinear contribution depends on both the system size and the shear rate.
In the small system-size regime ($L \lesssim 32$), the viscosity correction $\Delta \eta$ exhibits a pure logarithmic growth, $\sim \log L$.
Crucially, in this regime, the data points for different shear rates collapse onto a single curve.
This behavior indicates that the macroscopic transport remains in the near-equilibrium regime.

As the system size increases, however, the growth of $\Delta \eta$ deviates from this logarithmic trend and eventually saturates to a constant value.
This saturation explicitly marks the boundary where the near-equilibrium RG prediction breaks down.
As is evident from the figure, this breakdown depends on both the system size and the shear rate; the shear-induced suppression effect becomes increasingly prominent for larger system sizes and higher shear rates.

\subsection{Classification based on the Reynolds Number}
\label{sec6-2}
To theoretically extract the dependencies of $\eta_{\rm obs}$ on the system size and the shear rate, we approximate the discrete sum in Eq.~\eqref{eq:eta_oneloop} as a continuous integral and substitute the Lutsko-Dufty expressions [Eqs.~\eqref{eq:CTT_theory} and \eqref{eq:CLL_theory}] for the correlation functions.
The key to evaluating this integral lies in the competition between two scales: the finite-size cutoff ($2\pi/L$) and the shear-induced crossover wavenumbers [$k_c^{T,L}$, Eqs.~\eqref{eq:k_cross_T} and \eqref{eq:k_cross_L}], which can be classified based on the Reynolds number $Re \equiv \dot{\gamma}L^2/\nu_0$~\cite{Ortiz_de_Zarate2019-pp}.

\paragraph{(i) Low-Reynolds-Number Regime ($Re \ll 1$):}
The low-Reynolds-number condition, $Re \equiv \dot{\gamma}L^2/\nu_0 \ll 1$, is equivalently expressed as $2\pi/L \gg k_c^T, k_c^L$.
Under this condition, $C_{LL}^{\rm lin}(\bm{k})$ and $C_{TT}^{\rm lin}(\bm{k})$ reduce to the viscous-dominated asymptotic forms in Eq.~\eqref{eq:ld_viscous_scaling_main}.
Evaluating the integral using these asymptotic forms yields:
\begin{align}
\eta_{\rm obs} &\approx \eta_0 - \frac{\rho_0}{\dot{\gamma}} \int_{2\pi/L}^{2\pi/a_{\rm uv}} \frac{d^2\bm{k}}{(2\pi)^2} \hat{k}_x \hat{k}_y C_{LL}^{\rm lin}(\bm{k}) + \frac{\rho_0}{\dot{\gamma}} \int_{2\pi/L}^{2\pi/a_{\rm uv}} \frac{d^2\bm{k}}{(2\pi)^2} \hat{k}_x \hat{k}_y C_{TT}^{\rm lin}(\bm{k}) \nonumber \\
&\approx \eta_0 + \frac{k_B T}{\Gamma_0} \int_{2\pi/L}^{2\pi/a_{\rm uv}} \frac{d^2\bm{k}}{(2\pi)^2} \frac{k_x^2k_y^2}{k^6} + \frac{k_B T}{\nu_0} \int_{2\pi/L}^{2\pi/a_{\rm uv}} \frac{d^2\bm{k}}{(2\pi)^2} \frac{k_x^2k_y^2}{k^6} \nonumber \\
&= \eta_0 + \frac{k_B T \rho_0}{16\pi}\left(\frac{1}{\eta_0 + \zeta_0} + \frac{1}{\eta_0}\right) \log \left(\frac{L}{a_{\rm uv}}\right).
\label{eq:etaobs-one-loop-equilibrium}
\end{align}
This demonstrates that for sufficiently small Reynolds numbers, the observed viscosity becomes independent of the shear rate $\dot{\gamma}$ and exhibits a pure logarithmic divergence with the system size $L$.
This limiting form perfectly coincides with the near-equilibrium RG prediction in the weak-renormalization regime [Eq.~\eqref{eq:obs_eta_pert}].
This equivalence not only demonstrates that the one-loop RG approximation reduces to the leading-order perturbative result when renormalization effects are small, but also verifies that the near-equilibrium RG prediction is valid in the low-Reynolds-number regime.

\paragraph{(ii) Large-Reynolds-Number Regime ($Re \gg 1$):}
The large-Reynolds-number condition, $Re \equiv \dot{\gamma}L^2/\nu_0 \gg 1$, is equivalently expressed as $2\pi/L \ll k_c^T, k_c^L$.
Under this condition, the accessible wavenumbers in the system extend below the crossover wavenumbers $k_c^{T,L}$.
However, as shown by the shear-dominated asymptotic forms in Eq.~\eqref{eq:ld_shear_scaling_main} and discussed in Appendix~\ref{secC}, the contributions to the viscosity correction from the low-wavenumber region ($k < k_c^{T,L}$) are suppressed relative to the $k^{-2}$ behavior extrapolated from the viscous-dominated regime.
Consequently, the dominant contribution to the integral comes from the viscous-dominated region ($k > k_c^{T,L}$).
Thus, we evaluate the integral using the viscous-dominated asymptotic forms:
\begin{align}
    \eta_{\rm obs} &\approx \eta_0 - \frac{\rho_0}{\dot{\gamma}} \int_{2\pi/L}^{2\pi/a_{\rm uv}} \frac{d^2\bm{k}}{(2\pi)^2} \hat{k}_x \hat{k}_y C_{LL}^{\rm lin}(\bm{k}) + \frac{\rho_0}{\dot{\gamma}} \int_{2\pi/L}^{2\pi/a_{\rm uv}} \frac{d^2\bm{k}}{(2\pi)^2} \hat{k}_x \hat{k}_y C_{TT}^{\rm lin}(\bm{k}) \nonumber \\
    &\approx \eta_0 - \frac{\rho_0}{\dot{\gamma}} \int_{k_c^L}^{2\pi/a_{\rm uv}} \frac{d^2\bm{k}}{(2\pi)^2} \hat{k}_x \hat{k}_y C_{LL}^{\rm lin}(\bm{k}) + \frac{\rho_0}{\dot{\gamma}} \int_{k_c^T}^{2\pi/a_{\rm uv}} \frac{d^2\bm{k}}{(2\pi)^2} \hat{k}_x \hat{k}_y C_{TT}^{\rm lin}(\bm{k}) \nonumber \\
    &= \eta_0 + \frac{k_B T\rho_0}{16\pi}\left[\frac{1}{\eta_0 + \zeta_0} \log \left(\frac{k_{\rm uv}}{k_c^L}\right) + \frac{1}{\eta_0} \log \left(\frac{k_{\rm uv}}{k_c^T}\right) \right] \nonumber \\
    &\simeq \eta_0 + \frac{k_B T\rho_0}{16\pi}\left[\frac{1}{\eta_0 + \zeta_0} \log \left(k_{\rm uv}\sqrt{\frac{\eta_0+\zeta_0}{\rho_0\dot{\gamma}}}\right) + \frac{1}{\eta_0} \log \left(k_{\rm uv}\sqrt{\frac{\eta_0}{\rho_0\dot{\gamma}}}\right) \right].
    \label{eq:etaobs-one-loop-shear}
\end{align}
Here, $k_{\rm uv} = 2\pi/a_{\rm uv}$.
This demonstrates that for sufficiently large Reynolds numbers, the observed viscosity becomes independent of the system size $L$ and exhibits a logarithmic dependence on the shear rate $\dot{\gamma}$~\cite{Onuki1979-ji, Ernst1978-yq}.

\section{Concluding Remarks}
\label{sec7}
In this study, we performed DNS of the fluctuating NS equations to investigate nonequilibrium long-range correlations and anomalous transport in two-dimensional sheared fluids.
To achieve this, we developed a highly accurate numerical scheme that incorporates shear-periodic boundary conditions.
By combining an operator-splitting method with discrete Fourier interpolation, our solver successfully captures the bulk properties of the sheared fluid without introducing numerical dissipation or artificial wall confinement effects.

Using this solver, we first performed a direct quantitative test of the Lutsko-Dufty theory for static velocity correlations~\cite{Lutsko1985-zb}.
The linearized DNS validate the mode-decoupling approximation [Eq.~\eqref{eq:CLT_zero}] and agree quantitatively with the analytical expressions [Eqs.~\eqref{eq:CTT_theory} and \eqref{eq:CLL_theory}], including the predicted asymptotic scaling and amplitudes, across the crossover from the viscous-dominated short-wavelength regime to the shear-dominated long-wavelength regime (Figs.~\ref{fig3} and \ref{fig4}).
Simulations of the full nonlinear fluctuating NS equations further show that the Lutsko-Dufty predictions remain quantitatively accurate when nonlinear effects are sufficiently weak (Fig.~\ref{fig5}).
These results demonstrate the quantitative validity of the Lutsko-Dufty predictions within fluctuating hydrodynamics.

Our validation of the Lutsko-Dufty theory provides a useful reference for interpreting microscopic simulations, where low-wavenumber deviations from the Lutsko-Dufty predictions have been reported~\cite{Nakano2022-kv}.
These deviations may reflect a departure from the linear hydrodynamic regime, finite-size effects, or microscopic physics beyond the fluctuating-hydrodynamic description.
Clarifying their origin is left for future work.

Moving beyond the linear regime, we simulated the full nonlinear fluctuating NS equations to quantitatively assess the one-loop RG prediction.
By measuring the observed viscosity (Fig.~\ref{fig6}), we found that the one-loop RG prediction remains quantitatively accurate up to a relative renormalization correction of $\Delta \eta / \eta_0 \approx 1$, well beyond the range of validity of the lowest-order perturbative prediction.
Furthermore, we identified the range over which the near-equilibrium one-loop RG prediction can be applied under uniform shear (Fig.~\ref{fig7}).
Its applicability is controlled by the Reynolds number: the prediction remains valid in the low-Reynolds-number regime, whereas shear-induced suppression of long-wavelength fluctuations becomes significant in the large-Reynolds-number regime.

Although standard macroscopic hydrodynamics has long demonstrated quantitative predictive capability, comparable tests of fluctuating hydrodynamics have remained challenging because of the limitations of particle-based simulations and the approximations required in analytical treatments.
Our work demonstrates that DNS provides a direct route to such quantitative tests.
We expect that this approach will facilitate further quantitative assessments of predictions from fluctuating hydrodynamics across a broader range of nonequilibrium systems.

Finally, we comment on the numerical methodology employed in this paper.
The combination of the operator-splitting method and Fourier interpolation, used here to extract the bulk properties under shear, offers high precision despite its ease of implementation.
Therefore, we expect its broader application to various phenomena occurring under uniform shear flow, such as phase transitions~\cite{Winter2010-mp, Nakano2021-bt, Saracco2021-jk}.

\FloatBarrier
\appendix
\section{Averaging procedure}
\label{secA}

In this appendix, we summarize the numerical procedures used for measuring physical quantities in our simulations.
To ensure that the system reaches a steady state, we first perform a relaxation run for a sufficiently large number of steps.
Following this, we perform an observation run.
During the observation period, the static correlation functions were measured every $10^5$ steps, and the observed viscosity was calculated every $10^3$ steps.
The above procedure is repeated for multiple independent simulations with different noise realizations.
The number of steps for relaxation and averaging and the number of samples are summarized in Table~\ref{tab2}.

\FloatBarrier
\begin{table}[tb]
\centering
\small
\begin{tabularx}{\columnwidth}{@{\extracolsep{\fill}}c|c|c|c|c|c|c|c}
\hline
Target figures & $N$ & $\eta_0$ & $\zeta_0/\eta_0$ & $\dot{\gamma}$ & \makecell{Relaxation\\steps} & \makecell{Averaging\\steps} & Samples \\
\hline
Figs.~\ref{fig3}, \ref{fig4}, and \ref{fig8} & $768$ & $0.1$ & $1$ & $0.1$ & $1.0\times 10^7$ & $1.0\times 10^7$ & $72$ \\
Fig.~\ref{fig5} & $768$ & $1.0$ & $1$ & $0.1$ & $1.0\times 10^7$ & $1.0\times 10^7$ & $72$ \\
Figs.~\ref{fig6} and \ref{fig10} & $96$ & $0.12$--$4.0$ & $4$ & $0.02$ & $0.5\times 10^7$ & $2.5\times 10^7$ & $2304$ \\
Fig.~\ref{fig7} & $48$ & $1.0$ & $1$ & $0.01$--$0.05$ & $0.5\times 10^7$ & $2.5\times 10^7$ & $2304$ \\
Fig.~\ref{fig7} & $96$ & $1.0$ & $1$ & $0.01$--$0.05$ & $0.5\times 10^7$ & $2.5\times 10^7$ & $2304$ \\
Fig.~\ref{fig7} & $192$ & $1.0$ & $1$ & $0.01$--$0.05$ & $0.5\times 10^7$ & $2.5\times 10^7$ & $576$ \\
Fig.~\ref{fig7} & $384$ & $1.0$ & $1$ & $0.01$--$0.05$ & $0.5\times 10^7$ & $2.5\times 10^7$ & $288$ \\
Fig.~\ref{fig7} & $768$ & $1.0$ & $1$ & $0.01$--$0.05$ & $0.5\times 10^7$ & $2.5\times 10^7$ & $144$ \\
Figs.~\ref{fig9} and \ref{fig10} & $96$ & $0.12$--$4.0$ & $1, 19$ & $0.02$ & $0.5\times 10^7$ & $2.5\times 10^7$ & $2304$ \\
\hline
\end{tabularx}
\caption
{
Simulation parameters, numbers of time steps used for relaxation and averaging, and numbers of independent samples.
Physical times are obtained by multiplying the number of steps by $\Delta t=0.0001$.
}
\label{tab2}
\end{table}

\section{Two-dimensional heat maps of the velocity correlations}
\label{secB}

\begin{figure}[!t]
\centering
\includegraphics{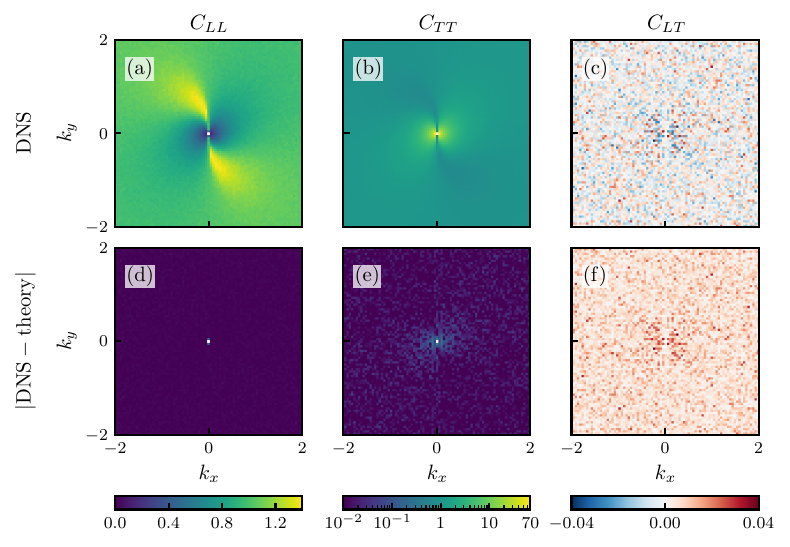}
\caption{\label{fig8}
Two-dimensional heat maps of the static velocity correlations in the linearized fluctuating NS equations.
The upper row shows the DNS data, and the lower row shows the absolute pointwise difference between the DNS and the Lutsko-Dufty theoretical expressions, $|C_{\rm DNS}(\bm{k})-C_{\rm theory}(\bm{k})|$.
The columns correspond to the longitudinal correlation $C_{LL}(\bm{k})$, the transverse correlation $C_{TT}(\bm{k})$, and the cross-correlation $C_{LT}(\bm{k})$, respectively.
Within each column, the same color scale is used for the DNS result and the corresponding absolute difference; a logarithmic color scale is used for $C_{TT}$.
The displayed range is $-2\leq k_x,k_y\leq2$, and the zero mode is omitted.
The parameters are the same as in Fig.~\ref{fig3}.
}
\end{figure}

Figure~\ref{fig8} presents two-dimensional heat maps of the velocity correlations whose one-dimensional cuts are shown in Fig.~\ref{fig3}.
Panels (a)--(c) show the DNS results for $C_{LL}(\bm{k})$, $C_{TT}(\bm{k})$, and $C_{LT}(\bm{k})$, respectively, while panels (d)--(f) show the corresponding absolute differences from the Lutsko-Dufty expressions, $|C_{\rm DNS}(\bm{k})-C_{\rm theory}(\bm{k})|$.
These differences remain small throughout the plotted wave-number plane, confirming that the quantitative agreement extends beyond the one-dimensional cuts shown in Fig.~\ref{fig3}.

\section{Derivation of Eqs.~\eqref{eq:k_cross_T}--\eqref{eq:ld_shear_scaling_main}}
\label{secC}

In this appendix, we derive the crossover wavenumbers and asymptotic forms summarized in Eqs.~\eqref{eq:k_cross_T}--\eqref{eq:ld_shear_scaling_main}.

\subsection{Asymptotic forms for generic wavevector directions}

For completeness, we reproduce here the Lutsko-Dufty expressions given in Eqs.~\eqref{eq:CTT_theory} and \eqref{eq:CLL_theory}:
\begin{align}
    C_{TT}(\bm{k})
    &=
    \frac{k_B T}{\rho_0}
    +
    \dot{\gamma}\frac{2k_B T}{\rho_0}
    \frac{k_x}{k^2}
    \int_0^\infty ds\,
    (k_y+\dot{\gamma}s k_x)
    \exp[-2\nu_0 I(\bm{k},s)],
    \label{eq:CTT_theory_app}
    \\
    C_{LL}(\bm{k})
    &=
    \frac{k_B T}{\rho_0}
    -
    \dot{\gamma}\frac{k_B T}{\rho_0}
    k k_x
    \int_0^\infty ds\,
    \frac{k_y+\dot{\gamma}s k_x}
    {[k_x^2+(k_y+\dot{\gamma}s k_x)^2]^{3/2}}
    \exp[-\Gamma_0 I(\bm{k},s)].
    \label{eq:CLL_theory_app}
\end{align}
Here, $\Gamma_0=(\eta_0+\zeta_0)/\rho_0$, and
\begin{align}
    I(\bm{k},s)
    &=
    sk^2+\dot{\gamma}s^2k_xk_y
    +\frac{\dot{\gamma}^2s^3k_x^2}{3}.
    \label{eq:ld_I_app}
\end{align}
These expressions involve a competition between viscous damping and shear advection.
The balance between these effects defines the transverse and longitudinal crossover wavenumbers as
\begin{align}
    k_c^T \sim \sqrt{\frac{\dot{\gamma}}{\nu_0}},
    \qquad
    k_c^L \sim \sqrt{\frac{\dot{\gamma}}{\Gamma_0}}.
    \label{eq:ld_cross_app}
\end{align}
These wavenumbers separate the viscous-dominated and shear-dominated asymptotic regimes of the Lutsko-Dufty expressions.
The corresponding asymptotic forms appear as Eqs.~\eqref{eq:ld_viscous_scaling_main} and \eqref{eq:ld_shear_scaling_main}, respectively, in the main text.

We begin with the viscous-dominated regime.
For the transverse and longitudinal modes, this regime corresponds to
$k\gg k_c^T$ and $k\gg k_c^L$, respectively.
In this regime, the shear-advection terms in $I(\bm{k},s)$ are perturbatively small compared with the diffusive term $sk^2$.
Indeed, for the transverse mode, the exponential factor in Eq.~\eqref{eq:CTT_theory_app} is mainly controlled by the time scale
$s_T\sim (2\nu_0 k^2)^{-1}$.
At this time scale, the three terms in Eq.~\eqref{eq:ld_I_app} are estimated as
\begin{align}
    k^2s_T
    &\sim
    \frac{1}{2\nu_0},
    \nonumber \\
    \dot{\gamma}k_xk_y s_T^2
    &\sim
    \frac{1}{4\nu_0}
    \frac{\dot{\gamma}}{\nu_0 k^2}
    \hat{k}_x\hat{k}_y,
    \nonumber \\
    \frac{\dot{\gamma}^2k_x^2}{3}s_T^3
    &\sim
    \frac{1}{24\nu_0}
    \left(
    \frac{\dot{\gamma}}{\nu_0 k^2}
    \right)^2
    \hat{k}_x^2 .
\end{align}
Thus, under the condition $k\gg k_c^T\sim\sqrt{\dot{\gamma}/\nu_0}$, the latter two terms are subleading, and we may approximate
\begin{align}
    I(\bm{k},s)
    \simeq
    k^2s.
    \label{eq:ld_diffusive_I}
\end{align}
In the same approximation, $k_y+\dot{\gamma}s k_x\simeq k_y$.
Substituting these approximations into Eq.~\eqref{eq:CTT_theory_app}, we obtain
\begin{align}
    C_{TT}^{\rm lin}(\bm{k})
    &\simeq
    \frac{k_B T}{\rho_0}
    + \dot{\gamma}\,\frac{2k_BT}{\rho_0}\frac{k_xk_y}{k^2}
    \int_0^\infty ds\, e^{-2\nu_0 k^2s}
    \nonumber \\
    &=
    \frac{k_B T}{\rho_0}
    + \dot{\gamma}\,\frac{k_BT}{\eta_0}\frac{k_xk_y}{k^4}.
    \label{eq:ld_diffusive_T_general}
\end{align}
Similarly, for the longitudinal mode, the relevant decay time is
$s_L\sim (\Gamma_0 k^2)^{-1}$.
Repeating the same estimate under the condition $k\gg k_c^L\sim\sqrt{\dot{\gamma}/\Gamma_0}$, we obtain
\begin{align}
    C_{LL}^{\rm lin}(\bm{k})
    &\simeq
    \frac{k_B T}{\rho_0}
    - \dot{\gamma}\,\frac{k_BT}{\rho_0}\frac{k_xk_y}{k^2}
    \int_0^\infty ds\, e^{-\Gamma_0 k^2s}
    \nonumber \\
    &=
    \frac{k_B T}{\rho_0}
    - \dot{\gamma}\,\frac{k_BT}{\eta_0+\zeta_0}\frac{k_xk_y}{k^4}.
    \label{eq:ld_diffusive_L_general}
\end{align}
Equations~\eqref{eq:ld_diffusive_T_general} and \eqref{eq:ld_diffusive_L_general} show that the leading nonequilibrium correction has the angular dependence
$k_xk_y/k^2=\hat{k}_x\hat{k}_y$ and the characteristic $k^{-2}$ growth.
Using $\nu_0=\eta_0/\rho_0$ and $\Gamma_0=(\eta_0+\zeta_0)/\rho_0$, these equations give the viscous-dominated forms summarized in Eq.~\eqref{eq:ld_viscous_scaling_main} in the main text.

We next consider the shear-dominated regime, $k\ll k_c^T$ or $k\ll k_c^L$.
For the transverse mode, the exponential factor in Eq.~\eqref{eq:CTT_theory_app} is controlled by the time scale
\begin{align}
    s_T^{\rm sh}\sim
    \left(\nu_0\dot{\gamma}^2k_x^2\right)^{-1/3},
\end{align}
which is determined by the cubic term in $I(\bm{k},s)$.
At this time scale, the three terms in $I(\bm{k},s)$ are estimated as
\begin{align}
    k^2s_T^{\rm sh}
    &\sim
    \frac{1}{\nu_0}
    \left(\frac{\nu_0 k^2}{\dot{\gamma}}\right)^{2/3}
    |\hat{k}_x|^{-2/3},
    \nonumber \\
    \dot{\gamma} k_xk_y (s_T^{\rm sh})^2
    &\sim
    \frac{1}{\nu_0}
    \left(\frac{\nu_0 k^2}{\dot{\gamma}}\right)^{1/3}
    \frac{\hat{k}_x\hat{k}_y}{|\hat{k}_x|^{4/3}},
    \nonumber \\
    \frac{\dot{\gamma}^2k_x^2}{3}(s_T^{\rm sh})^3
    &\sim
    \frac{1}{\nu_0}.
\end{align}
Thus, under the condition $k\ll k_c^T\sim\sqrt{\dot{\gamma}/\nu_0}$, the cubic term dominates the integral in Eq.~\eqref{eq:CTT_theory_app}, and we may approximate
\begin{align}
    I(\bm{k},s)
    \simeq
    \frac{\dot{\gamma}^2k_x^2}{3}s^3.
    \label{eq:ld_shear_I_cubic}
\end{align}
The case $k_x=0$ is exceptional, because the shear-induced stretching vanishes.
Substituting Eq.~\eqref{eq:ld_shear_I_cubic} into Eq.~\eqref{eq:CTT_theory_app} gives
\begin{align}
    C_{TT}^{\rm lin}(\bm{k})-\frac{k_BT}{\rho_0}
    &\simeq
    \dot{\gamma}\frac{2k_BT}{\rho_0}\frac{k_x}{k^2}
    \int_0^\infty ds\,(k_y+\dot{\gamma} s k_x)
    \exp\left[-\frac{2\nu_0\dot{\gamma}^2k_x^2}{3}s^3\right].
\end{align}
Evaluating the integrals yields
\begin{align}
    C_{TT}^{\rm lin}(\bm{k})-\frac{k_BT}{\rho_0}
    &\simeq
    \frac{k_BT}{\rho_0}
    \left(\frac{2}{3}\right)^{1/3}
    \Gamma\left(\frac{2}{3}\right)
    \left(\frac{\dot{\gamma}}{\nu_0}\right)^{2/3}
    \frac{|k_x|^{2/3}}{k^2}
    \nonumber \\
    &\quad
    +\frac{k_BT}{\rho_0}
    \left(\frac{2}{3}\right)^{2/3}
    \Gamma\left(\frac{1}{3}\right)
    \left(\frac{\dot{\gamma}}{\nu_0}\right)^{1/3}
    \frac{k_xk_y}{k^2|k_x|^{2/3}}.
    \label{eq:CTT_shear_general}
\end{align}
Thus, the leading transverse contribution scales as $k^{-4/3}$, with an angle-dependent amplitude.
The leading term in Eq.~\eqref{eq:CTT_shear_general}, with $(2/3)^{1/3}\Gamma(2/3)\simeq1.183$, is the transverse shear-dominated form summarized in Eq.~\eqref{eq:ld_shear_scaling_main} in the main text.

Similarly, for the longitudinal mode, under the condition $k\ll k_c^L\sim\sqrt{\dot{\gamma}/\Gamma_0}$, we approximate
\begin{align}
    I(\bm{k},s)
    \simeq
    \frac{\dot{\gamma}^2k_x^2}{3}s^3.
    \label{eq:ld_shear_I_cubic_L}
\end{align}
To rewrite Eq.~\eqref{eq:CLL_theory_app}, we introduce the shear-advected wavevector
\begin{align}
    K(s)
    &\equiv
    |\bm{K}(s)|,
    \qquad
    \bm{K}(s)=(k_x,k_y+\dot{\gamma} s k_x).
    \label{eq:ld_Ks_def}
\end{align}
Under this approximation, Eq.~\eqref{eq:CLL_theory_app} becomes
\begin{align}
    C_{LL}^{\rm lin}(\bm{k})
    &\simeq
    \frac{k_BT}{\rho_0}
    \left[
    1-\dot{\gamma} kk_x
    \int_0^\infty ds\,
    \frac{k_y+\dot{\gamma} s k_x}{K(s)^3}
    \exp\left[-\frac{\Gamma_0\dot{\gamma}^2k_x^2}{3}s^3\right]
    \right].
\end{align}
The equilibrium term is exactly canceled by the part obtained by setting the exponential factor to unity:
\begin{align}
    \dot{\gamma} kk_x
    \int_0^\infty ds\,
    \frac{k_y+\dot{\gamma} s k_x}{K(s)^3}
    =1.
\end{align}
Therefore,
\begin{align}
    C_{LL}^{\rm lin}(\bm{k})
    &\simeq
    \frac{k_BT}{\rho_0}
    \dot{\gamma} kk_x
    \int_0^\infty ds\,
    \frac{k_y+\dot{\gamma} s k_x}{K(s)^3}
    \left[1-\exp\left(-\frac{\Gamma_0\dot{\gamma}^2k_x^2}{3}s^3\right)\right].
\end{align}
In the shear-dominated regime, the prefactor is approximated by its large-$s$ form:
\begin{align}
    \dot{\gamma} kk_x
    \frac{k_y+\dot{\gamma} s k_x}{K(s)^3}
    \simeq
    \frac{k}{\dot{\gamma}|k_x|}\frac{1}{s^2}.
\end{align}
Using this prefactor approximation, we find
\begin{align}
    C_{LL}^{\rm lin}(\bm{k})
    &\simeq
    \frac{k_BT}{\rho_0}
    \frac{k}{\dot{\gamma}|k_x|}
    \int_0^\infty ds\,\frac{1}{s^2}
    \left[1-\exp\left(-\frac{\Gamma_0\dot{\gamma}^2k_x^2}{3}s^3\right)\right]
    \nonumber \\
    &=
    \frac{k_BT}{\rho_0}
    3^{-1/3}\Gamma\left(\frac{2}{3}\right)
    \left(\frac{\dot{\gamma}}{\Gamma_0}\right)^{-1/3}
    \frac{k}{|k_x|^{1/3}}.
    \label{eq:CLL_shear_general}
\end{align}
Thus, the longitudinal correlation itself scales as $k^{2/3}$, again with an angle-dependent amplitude.
Using $3^{-1/3}\Gamma(2/3)\simeq0.939$, Eq.~\eqref{eq:CLL_shear_general} gives the longitudinal shear-dominated form summarized in Eq.~\eqref{eq:ld_shear_scaling_main} in the main text.

\subsection{Shear-induced suppression of long-range correlations}

The asymptotic forms derived above show how nonequilibrium long-range correlations are suppressed by uniform shear.
In the viscous-dominated regime, Eq.~\eqref{eq:ld_viscous_scaling_main} in the main text shows that the nonequilibrium parts of the correlations behave as
\begin{align}
    C_{TT}^{\rm lin}(\bm{k})-\frac{k_BT}{\rho_0}
    &\simeq
    \frac{k_BT}{\rho_0}
    \frac{\dot{\gamma}}{\nu_0 k^2}
    \hat{k}_x\hat{k}_y,
    \nonumber \\
    C_{LL}^{\rm lin}(\bm{k})-\frac{k_BT}{\rho_0}
    &\simeq
    -
    \frac{k_BT}{\rho_0}
    \frac{\dot{\gamma}}{\Gamma_0 k^2}
    \hat{k}_x\hat{k}_y .
    \label{eq:ld_diffusive_scaling_summary}
\end{align}
Thus, apart from the angular factor, both correlations exhibit the characteristic $k^{-2}$ growth.
This is the well-known long-range behavior of velocity correlations under weak shear.

However, this $k^{-2}$ growth cannot be extrapolated to arbitrarily small wavenumbers.
Below the crossover wavenumber $k_c^T$, the transverse relation in Eq.~\eqref{eq:ld_shear_scaling_main} gives, for a fixed generic direction with $k_x\neq0$,
\begin{align}
    C_{TT}^{\rm lin}(\bm{k})-\frac{k_BT}{\rho_0}
    &\sim
    \frac{k_BT}{\rho_0}
    \left(\frac{\dot{\gamma}}{\nu_0}\right)^{2/3}
    |\hat{k}_x|^{2/3}
    k^{-4/3}
    +O(k^{-2/3}).
    \label{eq:ld_shear_T_scaling_summary}
\end{align}
Therefore, the transverse correlation still grows toward small $k$, but the divergence is weakened from $k^{-2}$ to $k^{-4/3}$.

The suppression is even stronger for the longitudinal mode.
Below the crossover wavenumber $k_c^L$, the longitudinal relation in Eq.~\eqref{eq:ld_shear_scaling_main} gives
\begin{align}
    C_{LL}^{\rm lin}(\bm{k})
    &\sim
    \frac{k_BT}{\rho_0}
    \left(\frac{\dot{\gamma}}{\Gamma_0}\right)^{-1/3}
    |\hat{k}_x|^{-1/3}
    k^{2/3}.
    \label{eq:ld_shear_L_scaling_summary}
\end{align}
Thus, the longitudinal correlation itself vanishes as $k \to 0$, instead of developing a long-range enhancement.
These results show that uniform shear suppresses the $k^{-2}$ long-range growth of the velocity correlations.

\section{Supplementary Details for the RG Analysis in Sec.~\ref{sec5}}
\label{secD}

\subsection{Velocity correlations in the one-loop effective equations}
\label{secD1}
In Sec.~\ref{sec5}, the nonlinear fluctuating NS equations are treated within the one-loop RG approximation, yielding effective linearized equations with scale-dependent viscosities.
The corresponding steady-state correlations can be calculated by the same procedure.
Defining
\begin{align}
    K(s)
    \equiv
    |\bm{K}(s)|,
    \qquad
    \bm{K}(s)=(k_x,k_y+\dot{\gamma}s k_x),
\end{align}
and
\begin{align}
    \mathcal{I}_T(\bm{k},s)
    &=
    \int_0^s du\,
    \frac{\eta_{\rm ren}(K(u))}{\rho_0}K(u)^2,
    \nonumber \\
    \mathcal{I}_L(\bm{k},s)
    &=
    \int_0^s du\,
    \frac{G_{\rm ren}(K(u))}{\rho_0}K(u)^2,
    \qquad
    G_{\rm ren}(k)=\eta_{\rm ren}(k)+\zeta_{\rm ren}(k),
\end{align}
the resulting one-loop expressions for the velocity correlations are
\begin{align}
    C_{TT}^{\rm eff}(\bm{k})
    &=
    \frac{k_BT}{\rho_0}
    +
    \dot{\gamma}\frac{2k_BT}{\rho_0}
    \frac{k_x}{k^2}
    \int_0^\infty ds\,
    (k_y+\dot{\gamma}s k_x)
    \exp[-2\mathcal{I}_T(\bm{k},s)],
    \label{eq:CTT_eff_full} \\
    C_{LL}^{\rm eff}(\bm{k})
    &=
    \frac{k_BT}{\rho_0}
    -
    \dot{\gamma}\frac{k_BT}{\rho_0}
    k k_x
    \int_0^\infty ds\,
    \frac{k_y+\dot{\gamma}s k_x}{K(s)^3}
    \exp[-\mathcal{I}_L(\bm{k},s)],
    \label{eq:CLL_eff_full} \\
    C_{LT}^{\rm eff}(\bm{k})
    &\simeq 0.
    \label{eq:CLT_eff_zero}
\end{align}
When \(\eta_{\rm ren}(k)\) and \(G_{\rm ren}(k)\) are replaced by the constant bare values \(\eta_0\) and \(\eta_0+\zeta_0\), Eqs.~\eqref{eq:CTT_eff_full} and \eqref{eq:CLL_eff_full} reduce to the Lutsko-Dufty expressions in Eqs.~\eqref{eq:CTT_theory} and \eqref{eq:CLL_theory}.

\subsection{Viscous-dominated asymptotic forms of the one-loop correlations}
\label{secD2}

In the low-Reynolds-number regime, Eqs.~\eqref{eq:CTT_eff_full} and \eqref{eq:CLL_eff_full} further reduce to the viscous-dominated asymptotic forms.
Since the renormalized viscosities depend only on the magnitude $k=|\bm{k}|$, the leading angular structure derived in Appendix~\ref{secC} is unchanged.
Thus, in the viscous-dominated regime, the corresponding one-loop forms are obtained by the replacements
\begin{align}
    \eta_0 \to \eta_{\rm ren}(k),
    \qquad
    \eta_0+\zeta_0 \to G_{\rm ren}(k),
\end{align}
where $G_{\rm ren}(k)=\eta_{\rm ren}(k)+\zeta_{\rm ren}(k)$.
This gives
\begin{align}
    C_{TT}^{\rm eff}(\bm{k})
    &\simeq
    \frac{k_BT}{\rho_0}
    +\dot{\gamma}\,\frac{k_BT}{\eta_{\rm ren}(k)}\frac{k_xk_y}{k^4},
    \nonumber \\
    C_{LL}^{\rm eff}(\bm{k})
    &\simeq
    \frac{k_BT}{\rho_0}
    -\dot{\gamma}\,\frac{k_BT}{G_{\rm ren}(k)}\frac{k_xk_y}{k^4}.
    \label{eq:ld_diffusive_rg_general}
\end{align}
These are the asymptotic forms used in Sec.~\ref{sec5}.

\subsection{Observed viscosity in the low-Reynolds-number regime}
\label{secD3}
We derive Eq.~\eqref{eq:obs_eta_fns} in the low-Reynolds-number regime.
In this regime, we can use the viscous-dominated asymptotic forms in Eq.~\eqref{eq:ld_diffusive_rg_general}.
Substituting them into the expression for $\eta_{\rm obs}$ gives
\begin{align}
    \eta_{\rm obs}-\eta_0
    &=
    k_B T\rho_0
    \int_{k_{\rm IR}}^{k_{\rm uv}}
    \frac{d^2\bm{k}}{(2\pi)^2}
    \frac{k_x^2k_y^2}{k^6}
    \left[
        \frac{1}{\eta_{\rm ren}(k)}
        +
        \frac{1}{G_{\rm ren}(k)}
    \right],
\end{align}
where $G_{\rm ren}(k)=\eta_{\rm ren}(k)+\zeta_{\rm ren}(k)$, $k_{\rm uv}=2\pi/a_{\rm uv}$, and $k_{\rm IR}=2\pi/L$.
Writing $d^2\bm{k}=k\,dk\,d\theta$ and using
\begin{align}
    \int_0^{2\pi} d\theta\,\cos^2\theta\sin^2\theta = \frac{\pi}{4},
\end{align}
we obtain
\begin{align}
    \eta_{\rm obs}-\eta_0
    &=
    \frac{k_B T\rho_0}{16\pi}
    \int_{k_{\rm IR}}^{k_{\rm uv}}
    \frac{dk}{k}
    \left[
        \frac{1}{\eta_{\rm ren}(k)}
        +
        \frac{1}{G_{\rm ren}(k)}
    \right].
\end{align}
The one-loop RG flow equation for the shear viscosity, Eq.~\eqref{eq:chen_rg_eta}, reads
\begin{align}
    \frac{d\eta_{\rm ren}(k)}{d\log k}
    =
    -\frac{k_B T\rho_0}{16\pi}
    \left[
        \frac{1}{\eta_{\rm ren}(k)}
        +
        \frac{1}{G_{\rm ren}(k)}
    \right].
\end{align}
Comparing this flow equation with the preceding integral, we obtain
\begin{align}
    \eta_{\rm obs}-\eta_0
    &=
    -\int_{k_{\rm IR}}^{k_{\rm uv}}
    d\eta_{\rm ren}(k)
    =
    \eta_{\rm ren}(k_{\rm IR})-
    \eta_{\rm ren}(k_{\rm uv}).
\end{align}
Using $\eta_{\rm ren}(k_{\rm uv})=\eta_0$, this yields
\begin{align}
    \eta_{\rm obs}
    =
    \eta_{\rm ren}(k_{\rm IR})
    =
    \eta_{\rm ren}\left(k=\frac{2\pi}{L}\right).
\end{align}
Thus, the observed viscosity is the renormalized shear viscosity evaluated at the infrared cutoff, reproducing Eq.~\eqref{eq:obs_eta_fns}.

\begin{figure}
\centering
\includegraphics{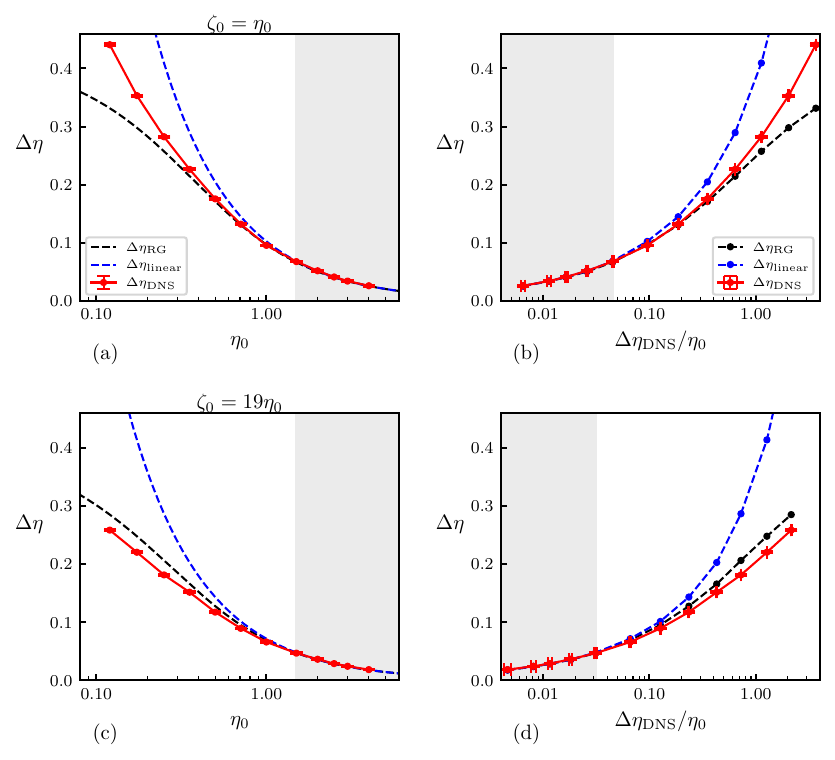}
\caption{\label{fig9}
Additional tests of the one-loop RG prediction at
$\zeta_0=\eta_0$ and $\zeta_0=19\eta_0$.
Panels (a) and (b) show the results for $\zeta_0=\eta_0$, while
panels (c) and (d) show those for $\zeta_0=19\eta_0$.
Panels (a) and (c) show the viscosity correction
$\Delta\eta=\eta_{\rm obs}-\eta_0$ as a function of $\eta_0$, and
panels (b) and (d) show the same results as functions of
$\Delta\eta_{\rm DNS}/\eta_0$.
The red symbols represent the DNS results, the black dashed lines represent
the one-loop RG prediction, and the blue dashed lines represent the
lowest-order perturbative prediction.
The theoretical curves are evaluated using
$a_{\rm uv}^{\rm eff}=0.518$ for $\zeta_0=\eta_0$ and
$a_{\rm uv}^{\rm eff}=0.517$ for $\zeta_0=19\eta_0$, determined
independently from the weak-renormalization data at each viscosity ratio.
The light-gray regions indicate the data range used to fit
$a_{\rm uv}^{\rm eff}$.
Other parameters are $N=96$, $h=1/6$, $\dot{\gamma}=0.02$,
$\rho_0=k_B T=1$, and $c_T^2=5000$.
}
\end{figure}
\section{Additional Tests for Sec.~\ref{sec5-3}}
\label{secE}

This appendix supplements the results in Sec.~\ref{sec5-3} by presenting additional simulations at $\zeta_0/\eta_0=1$ and $19$.
All numerical parameters other than the viscosity ratio $\zeta_0/\eta_0$ are the same as those used in Fig.~\ref{fig6}.
These additional data allow us to examine both the consistency of the effective UV cutoff and the quantitative validity of the one-loop RG prediction.

Following the procedure described in Sec.~\ref{sec5-3}, we independently determine $a_{\rm uv}^{\rm eff}$ for each $\zeta_0/\eta_0$ using only the data satisfying
\begin{align}
    \eta_0\geq1.5.
\end{align}
These data lie in the weak-renormalization regime,
$\Delta\eta_{\rm DNS}/\eta_0<0.1$.
The independently fitted values for the three datasets are
\begin{align}
    a_{\rm uv}^{\rm eff} =
    \begin{cases}
    0.518\pm0.022,
    \qquad {\rm for} \qquad \zeta_0/\eta_0=1,
    \\
    0.526\pm0.028,
    \qquad {\rm for} \qquad \zeta_0/\eta_0=4,
    \\
    0.517\pm0.031,
    \qquad {\rm for} \qquad \zeta_0/\eta_0=19.
    \end{cases}
\end{align}
The three estimates agree within approximately $2\%$.
This modest variation supports the interpretation that $a_{\rm uv}^{\rm eff}$ is determined primarily by the numerical lattice structure rather than by the fluid parameters.

In the weak-renormalization regime, the perturbative and RG predictions are nearly identical and both remain close to the DNS data.
As the normalized renormalization correction increases, the perturbative prediction increasingly deviates from the DNS, whereas the one-loop RG prediction describes the nonlinear growth of $\Delta\eta$ substantially more accurately.

The quantitative agreement at stronger renormalization depends somewhat on $\zeta_0/\eta_0$.
For $\zeta_0=4\eta_0$, the main-text results in Fig.~\ref{fig6} show that
the agreement persists up to the largest value examined,
$\Delta\eta_{\rm DNS}/\eta_0\simeq3$.
For $\zeta_0=\eta_0$ and $\zeta_0=19\eta_0$, the RG prediction also remains close to the DNS
when $\Delta\eta_{\rm DNS}/\eta_0$ exceeds unity.
The results for all three viscosity ratios show that the one-loop
RG prediction remains quantitatively reliable at least for
\begin{align}
    \frac{\Delta\eta_{\rm DNS}}{\eta_0}\lesssim1.
\end{align}
These supplementary results therefore support the validity criterion
stated in Sec.~\ref{sec5-3}.

\section{Incompressible Limit of the Viscosity Renormalization}
\label{secF}

In conventional deterministic hydrodynamics, the incompressible limit is usually associated with the low-Mach-number limit, which can be realized by taking $c_T\to\infty$ at fixed flow velocity.
By contrast, in fluctuating hydrodynamics, increasing the sound speed does not necessarily lead to the incompressible limit.
This can be understood from the linear analysis in Sec.~\ref{sec4-2} and Appendix~\ref{secC}.
The results obtained there are reproduced below.
The transverse and longitudinal correlations are given by
\begin{align}
    C_{TT}^{\rm lin}(\bm{k})
    &\simeq
    \frac{k_BT}{\rho_0}
    +\dot{\gamma}\frac{k_BT}{\rho_0\nu_0}
    \frac{k_xk_y}{k^4},
    \label{eq:incomp_diffusive_T}
    \\
    C_{LL}^{\rm lin}(\bm{k})
    &\simeq
    \frac{k_BT}{\rho_0}
    -\dot{\gamma}\frac{k_BT}{\rho_0\Gamma_0}
    \frac{k_xk_y}{k^4}
    \label{eq:incomp_diffusive_L}
\end{align}
in the diffusion-dominated regime ($k\gg k_c^T\sim\sqrt{\dot{\gamma}/\nu_0}$ for the transverse mode and $k\gg k_c^L\sim\sqrt{\dot{\gamma}/\Gamma_0}$ for the longitudinal mode), where $\nu_0=\eta_0/\rho_0$ and $\Gamma_0=(\eta_0+\zeta_0)/\rho_0$.
The correlations in the shear-dominated regime ($k\ll k_c^T$ for the transverse mode and $k\ll k_c^L$ for the longitudinal mode) are more strongly suppressed than those in the diffusion-dominated regime and are therefore not considered here.

These expressions show that the transverse and longitudinal correlations depend on $\nu_0$ and $\Gamma_0$, respectively, but not on $c_T$.
This independence shows that taking $c_T\to\infty$ does not, by itself, yield the incompressible limit.
Instead, the relative importance of the two modes is determined by the ratio of their damping coefficients:
\begin{align}
    \frac{\Gamma_0}{\nu_0}
    =
    1+\frac{\zeta_0}{\eta_0}.
\end{align}
Since the transverse contribution becomes dominant as $\zeta_0/\eta_0$ increases, the limit $\zeta_0/\eta_0\to\infty$ corresponds to the incompressible limit for nonequilibrium fluctuations.

The same conclusion follows directly from the compressible one-loop RG equations~\cite{Chen1995},
which are given by
\begin{align}
    \frac{d\eta_{\rm ren}(k)}{d\log k}
    &=
    -\frac{k_BT\rho_0}{16\pi}
    \left[
        \frac{1}{\eta_{\rm ren}(k)}
        +
        \frac{1}{G_{\rm ren}(k)}
    \right],
    \label{eq:chen_rg_eta_appF}
    \\
    \frac{dG_{\rm ren}(k)}{d\log k}
    &=
    -\frac{3k_BT\rho_0}{16\pi}
    \left[
        \frac{1}{\eta_{\rm ren}(k)}
        +
        \frac{1}{G_{\rm ren}(k)}
    \right],
    \label{eq:chen_rg_G_appF}
\end{align}
where
\begin{align}
    G_{\rm ren}(k)
    =
    \eta_{\rm ren}(k)+\zeta_{\rm ren}(k).
\end{align}
The terms proportional to $1/\eta_{\rm ren}(k)$ and
$1/G_{\rm ren}(k)$ in Eq.~\eqref{eq:chen_rg_eta_appF} represent the
transverse and longitudinal contributions, respectively.
At $k=k_{\rm uv}$, we impose
\begin{align}
    \eta_{\rm ren}(k_{\rm uv})=\eta_0,
    \qquad
    G_{\rm ren}(k_{\rm uv})=\eta_0+\zeta_0.
\end{align}
As $\zeta_0/\eta_0\to\infty$, $G_{\rm ren}(k)$ becomes much larger than
$\eta_{\rm ren}(k)$, and the longitudinal contribution
$1/G_{\rm ren}(k)$ vanishes.
Equation~\eqref{eq:chen_rg_eta_appF} then reduces to
\begin{align}
    \frac{d\eta_{\rm ren}(k)}{d\log k}
    =
    -\frac{k_BT\rho_0}{16\pi}
    \frac{1}{\eta_{\rm ren}(k)},
\end{align}
which is the incompressible RG equation derived by FNS for Model A.
Solving this equation between
$k_{\rm uv}=2\pi/a_{\rm uv}^{\rm eff}$ and
$k_{\rm IR}=2\pi/L$ gives the nonlinear incompressible prediction
\begin{align}
    \Delta\eta_{\rm FNS}
    =
    \sqrt{
        \eta_0^2
        +
        \frac{k_BT\rho_0}{8\pi}
        \log\left(\frac{L}{a_{\rm uv}^{\rm eff}}\right)
    }
    -\eta_0.
\end{align}

\begin{figure}[p]
\centering
\includegraphics[width=0.56\linewidth]{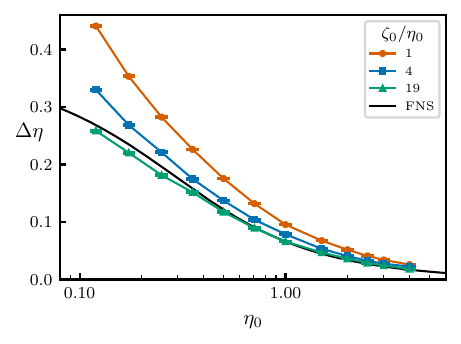}
\caption{\label{fig10}
Approach of the viscosity renormalization to the incompressible limit.
The colored markers connected by solid lines show the nonlinear DNS results
at $\zeta_0/\eta_0=1$, $4$, and $19$.
The black solid line is the incompressible FNS prediction for Model A, evaluated using
$a_{\rm uv}^{\rm eff}=0.517$.
Other parameters are $N=96$, $h=1/6$, $\dot{\gamma}=0.02$,
$\rho_0=k_B T=1$, and $c_T^2=5000$.
}
\end{figure}

Figure~\ref{fig10} compares the DNS results at
$\zeta_0/\eta_0=1$, $4$, and $19$ with the incompressible FNS limit.
As $\zeta_0/\eta_0$ increases, the DNS data approach the FNS curve.
In particular, the results at $\zeta_0=19\eta_0$ are already close to the
incompressible prediction over the range examined here.
This trend supports the theoretical discussion presented above.

\clearpage
\ack{
We thank K. Yokota, T. Tanogami, and R. Araki for valuable discussions.
HN is supported by JSPS KAKENHI Grant No. JP22K13978.
YM is supported by JSPS KAKENHI Grant No. JP25K07148 and the Ogawa science and technology foundation.
The numerical computation in this study has been done using the facilities of the Supercomputer Center, the Institute for Solid State Physics, the University of Tokyo. 
}





\bibliographystyle{apsrev4-2}
\bibliography{
    ref_fhd.bib,
    ref_fhd_theo.bib,
    ref_fhd_md.bib,
    ref_fhd_fns.bib,
    ref_fhd_lusduf.bib,
    ref_fhd_garcia.bib,
    ref_fhd_others.bib
}

\end{document}